\documentclass[aps,floats,prb,showpacs,twocolumn]{revtex4}

\newcommand{\kbar} {\mathchar'26\mkern-9muk}

\usepackage{amsfonts,amsmath} \usepackage{bm} \usepackage{dcolumn}
\usepackage{epsfig} \usepackage{latexsym}

\begin{document}

\title{Ehrenfest time in the weak dynamical localization}

\author{C. Tian$^{1}$, A. Kamenev$^{1}$, and A.  Larkin$^{1,2,3}$}

\affiliation{$^{1}$ Department of Physics, University of
Minnesota,
Minneapolis, MN 55455, USA\\
$^{2}$ William I. Fine Theoretical Physics Institute, University
of Minnesota,
Minneapolis, MN 55455, USA\\
$^{3}$ L. D. Landau Institute for Theoretical Physics, Moscow,
117940, Russia}

\date{\today}


\begin{abstract}
{\rm The quantum kicked rotor (QKR) is known to exhibit
    dynamical localization in the space of its angular momentum.
    The present paper is devoted to the systematic
    first--principal (without a regularizer)
    diagrammatic calculations of the weak--localization
    corrections for QKR. Our particular emphasis is on the
    Ehrenfest time regime -- the phenomena
    characteristic for the classical--to--quantum crossover of
    classically chaotic systems.}
\end{abstract}

\maketitle

\section{Introduction}
\label{Intro}

In recent years it became abundantly clear that driven quantum
systems exhibit behavior that is qualitatively different from
their classical counterparts. Indeed, a  driven classically
chaotic system exhibits Joel's heating. That is, its average
stored energy  increases at a steady rate.  In other words, such
behavior may be characterized as a diffusion in the system's phase
space. The remarkable feature of driven {\em quantum} systems is
finiteness of their phase space motion (localization). Such
dynamical localization phenomena was discussed in context of
pumped quantum dots \cite{Basko}, Bose-Einstein condensates
subject to pulses of optical standing  wave
\cite{Raizen95,Raizen99,Zhang04}, optical microcavity
\cite{Jacquod}, and other systems.

The simplest model that became a paradigm for studies of the
quantum dynamical localization is quantum kicked rotor (QKR). It
was numerical discovery of localization in QKR by Casati {\it et
al.} \cite{CCFI79,CIS81} in the late seventies that triggered the
broad interest  to the subject.    Recent progress  in  trapping
of cold atoms and optical manipulation with them lead to
experimental realization of the QKR with the unprecedented degree
of control \cite{Raizen95,Zhang04}. The kicked rotor is described
by the time--dependent Hamiltonian:
\begin{equation}
{\hat H}(t) =\frac{{\hat l}^2}{2}+ K \cos {\hat \theta} \sum_n
\delta \left(t-n\right)\, ,
                                           \label{Hamiltonian}
\end{equation}
where  angle $\hat\theta$ and angular momentum $\hat l$ are the
pair of canonically conjugated variables. The amplitude of the
kicks is described by the dimensionless parameter $K$, also known
as the classical stochastically parameter. It is the only
parameter of the corresponding classical problem. The quantum
problem possesses another dimensionless parameter: the effective
Planck constant $\kbar$. The latter enters the problem through the
canonical commutation relation: $[\hat\theta, {\hat l}]=i\kbar$.
The two parameters $K$ and $\kbar$ are  straightforwardly related
to the optical wavelength, amplitude and atomic mass in  cold
atoms experiments \cite{Raizen95,Raizen99,Ammann}.

Historically, the {\em  classical } kicked rotor, or standard
mapping, first introduced by Chirikov,  served as the prototype
model for various transport processes in plasmas
\cite{Chirikov79,LL}. It was established that the classical
dynamics of the kicked rotor exhibits complicated behavior. Most
notably it demonstrates the transition from the regular to chaotic
motion \cite{Chirikov79,LL}. In particular, for sufficiently large
classical parameter $K~(\gtrsim 5)$, the chaotic diffusion takes
place in the space of angular momentum \cite{Chirikov79,LL}. The
latter is associated with the unlimited diffusive expansion of an
initially narrow  momenta distribution: $\delta\langle
l^2(t)\rangle=2D_{cl}t\, .$
The classical diffusion constant, $D_{cl}$, was a subject of
numerous studies  \cite{Chirikov79,LL} and is well understood by
now \cite{Rec81,KFA00}. For large stochastically parameter $K\gg
1$, one finds $D_{cl}\approx K^2/4 +O(K^{3/2})$, where the omitted
corrections posses an oscillatory dependence on $K$.

The pioneering numerical studies of Casati {\it et al.}
\cite{CCFI79,CIS81} revealed that the corresponding {\em quantum}
system, $\kbar \neq 0$, behaves in a dramatically different way.
The initial diffusive expansion (that is heating) saturates after
a certain time, $t_L$, and the momentum distribution width
stabilizes at $\delta\langle l^2(t)\rangle \sim \xi^2 =
D_{cl}t_L$. It was soon suggested in Ref.~\onlinecite{FGP82} that
similarly to Anderson localization, quantum phase interference may
lead to the suppression of classical diffusion for long enough
time. This heuristic idea was complemented by mapping  QKR onto a
one-dimensional tight-binding Anderson model with the
pseudo--random potential \cite{FGP82,FGP84}. Such interpretation
leads to the estimate of the localization length as
$\xi=D_{cl}/\kbar$ and thus $t_L=D_{cl}/\kbar^2$. The similarity
was further confirmed by  studies of a perturbation that breaks
the ``time-reversal symmetry'' (TRS) of the QKR \cite{BS92}. Such
perturbation suppresses  the survival probability by a factor of
two. That is closely analogous to the doubling of the Anderson
localization length by the static magnetic field, destroying the
interference between a trajectory and its time-reversal partner
\cite{LR85,Efetov97}.

If the momentum localization length, $\xi$, is much longer than
the  ``microscopic'' scale of the angular momentum (that is
$\kbar$) then  $K\gg \kbar$  and thus $t_L\gg 1$. In this case
there is a parametrically long crossover regime from the classical
diffusion: $\delta\langle l^2(t)\rangle \approx 2D_{cl}t$ for $1
  <   t\ll t_L$ to the strong localization: $\delta\langle l^2(t)\rangle
=\xi^2$ for $t  >   t_L$. One may be able to develop a systematic
perturbation theory in powers of $(t/t_L)\ll 1$, analogous to the
weak--localization loop expansion in the Anderson localization
theory \cite{GLKh79}. Such task was undertaken by Altland
\cite{Altland93}, who found for the one--loop correction:
$\delta\langle l^2(t)\rangle = 2D_{cl}t\,(1-0.75 \sqrt{t/t_L}\,)$.
It was suggested furthermore that the universal long-time behavior
of the QKR is described by the diffusive supersymmetric nonlinear
$\sigma$-model \cite{AZ96} similar to those employed in the
localization theory \cite{Efetov97}.

Those calculations essentially map the  QKR on a quantum particle
in the field of a {\em white--noise} random potential. While such
analogy is reasonable at long time scales $t\sim t_L\gg 1$, it
fails to recognize details of the classical to quantum crossover
at intermediate time scales. Indeed, a quantized classically
chaotic system requires a certain time scale, called the Ehrenfest
time, $t_E$, to develop quantum interference effects. This fact
was realized independently in various contexts \cite{LO68,BZ78}
and now days is well documented in the literature
\cite{CIS81,BZ78,Izrailev90,Beenakker,Beenakker03,TTSB03,AL96}.
The physics behind this fact is as follows. To experience the
quantum interference, two classical trajectories must converge to
a region of the phase space of the size $\delta l \delta\theta
\approx \kbar$. Convergence (and divergence) of trajectories in a
classically chaotic system is governed by the Lyapunov instability
exponent $\lambda$ as e.g. $\delta\theta (t)\sim \exp\{-\lambda
t\}$. It thus takes time $t_E\sim \lambda^{-1}\ln(1/\kbar)$ before
the interference effects can reveal themselves.

As we show below, for $K\gg 1$ the quantitative definition of the
Ehrenfest time for the QKR is given by
\begin{equation}
t_E={1\over \lambda}\ln \sqrt{K\over \kbar}\, ,
                                  \label{ehrenfest}
\end{equation}
while the classical Lyapunov exponent is  $\lambda=\ln(K/2)$
\cite{Chirikov79}. Therefore there is the parametric regime $1\ll
K\ll \kbar^{-1}$, or more precisely:
\begin{equation}
\label{regime}
 \ln\left({1\over\kbar}\right)\gg \ln K  >  0 \, ,
\end{equation}
where exists  a wide separation of the relevant time scales:
\begin{equation}
\label{scales}
 1\ll t_E \ll t_L\, .
\end{equation}
One may  expect that such  regime is amenable for an  {\em
analytical} treatment of the classical--to--quantum crossover.
This problem was first tackled by Aleiner and Larkin \cite{AL96}
in the context of random classical (long-range) potential
scattering (e.g., random Lorentz gases). However, due to the
complexity of the Lorentz gas classical dynamics, their treatment
required a regularization. The latter is essentially a weak
quantum scattering potential added to the Lorentz gas.

The purpose of this paper is to develop a systematic
first--principle analytic treatment of the QKR. In particular, we
are able to
incorporate the semiclassical dynamics at the scale of $t_E$ into
the  weak dynamical localization theory without introducing any
regularization. The QKR allows thus to demonstrate explicitly an
essential point: {\em existence of the dynamical localization is
an intrinsic property of quantized classically chaotic systems --
not an artefact of an extraneous regularization}. (Remarkably, the
Ehrenfest time does {\it not} dependend on the regularizer
strength \cite{VL03,TKL04,MA04}.) This observation is fully
consistent with the early studies of Ehrenfest time
\cite{LO68,BZ78}, suggested the existence of such time scale. For
the time interval $t\lesssim t_L$, our approach fully encompasses
the Ehrenfest regime. The results were reported in the short
letter \cite{TKL04}. The main result for the time--dependent
spread of the wave--packet may be formulated as:
\begin{equation}
\delta \left\langle l^2(t)\right\rangle = 2D_{cl}\left[ t
-\frac{4}{3\sqrt\pi}\, \theta (t- 4t_E)\frac{\left(t-
4t_E\right)^{3/2}}{t_L^{1/2}} \right]\, ,
                                           \label{resultApp}
\end{equation}
where $\theta(t)$ is the standard Heaviside step--function. At
intermediate times, $t_E\ll t   <   t_L$, Eq.~(\ref{resultApp})
crosses over to the standard weak--localization correction
\cite{Altland93}. At shorter times, $t\approx t_E$, there is a
delay in developing localization given by $4t_E$. A few comments
are in due: (i) the actual delay is not absolutely sharp, as
suggested by Eq.~(\ref{resultApp}). There are  exponentially small
deviation from the straight line $2D_{cl} t$ even for $t  < 4t_E$,
which exact shape is calculated below. (ii) Eq.~(\ref{resultApp})
describes quantum correction {\em linear} in $\kbar$ that appears
to be delayed by $4t_E$. As first noticed by Shepelansky
\cite{She87}, {\em quadratic} in $\kbar$ corrections show up even
at earlier time. However, for at least three first kicks, they may
be fully absorbed into a renormalization of the diffusion constant
\cite{She87} ($\delta D_{cl}\sim \kbar^2$, essentially due to the
change in the scattering cross--section). It is thus an
oversimplification to claim the absence of quantum effects at $t <
4t_E$. (iii) Eq.~(\ref{resultApp}) constitutes the one--loop
weak--localization correction. Below we report also the results of
the two--loop calculation. It brings the next order correction
$\sim \theta(t-6t_E)(t-6t_E)^2/t_L$ that also ``protects'' the
early time evolution from the localization effects. It is still an
open problem to sum up the entire series to develop a theory of
strong localization that accounts for the Ehrenfest time
phenomena.

Most of the existing experiments on atomic gases
\cite{Raizen99,Raizen00,kbarexp} do not fall down in the
parametric regime (\ref{regime}), but rather have $\kbar\approx
1$. In this case $t_E\approx 1$ and our result,
Eq.~(\ref{resultApp}), can only be viewed as a qualitative one. We
discuss below other possible realizations of the QKR that utilizes
driven Josephson junctions \cite{MRBF04,IGSGZ02}. Such systems may
prove to be more suitable for exploring the parametric regime
(\ref{regime}) and thus for a quantitative comparison with the
theory.

The outline of the rest of this paper is as follows:
Sec.~\ref{GenCon} is devoted to a qualitative semiclassical
picture of the weak--dynamical localizations and the Lyapunov
regime. Sections~\ref{classicalDC}, \ref{WDLKR} and
\ref{QuantumCorr2Loop} serve to quantify these ideas. In
Sec.~\ref{classicalDC} the diffusion in the phase space of the
kicked rotor is obtained as a classical approximation to the full
quantum propagation. Sec.~\ref{WDLKR} is the central part of the
present work. It formulates a general framework to deal with the
weak dynamical localization at the semiclassical level. In
particular, we calculate the frequency-dependent one-loop
correction to the classical diffusion coefficient and study its
effect on the momentum dispersion. This formalism is applied in
Sec.~\ref{QuantumCorr2Loop} to a modified QKR with broken
time--reversal symmetry. The frequency-dependent quantum
corrections are calculated  at the two-loop level.  Experimental
realizations of  some driven quantum systems are discussed in
Sec.~\ref{EhrDri}. The effects of noise and dephasing are subject
of Sec.~\ref{noise}. We conclude in Sec.~\ref{CON}. Some technical
details are delegated to   Appendices.

\section{Qualitative considerations}
\label{GenCon}

\begin{figure}[h]
  \centerline{\epsfxsize=3in\epsfbox{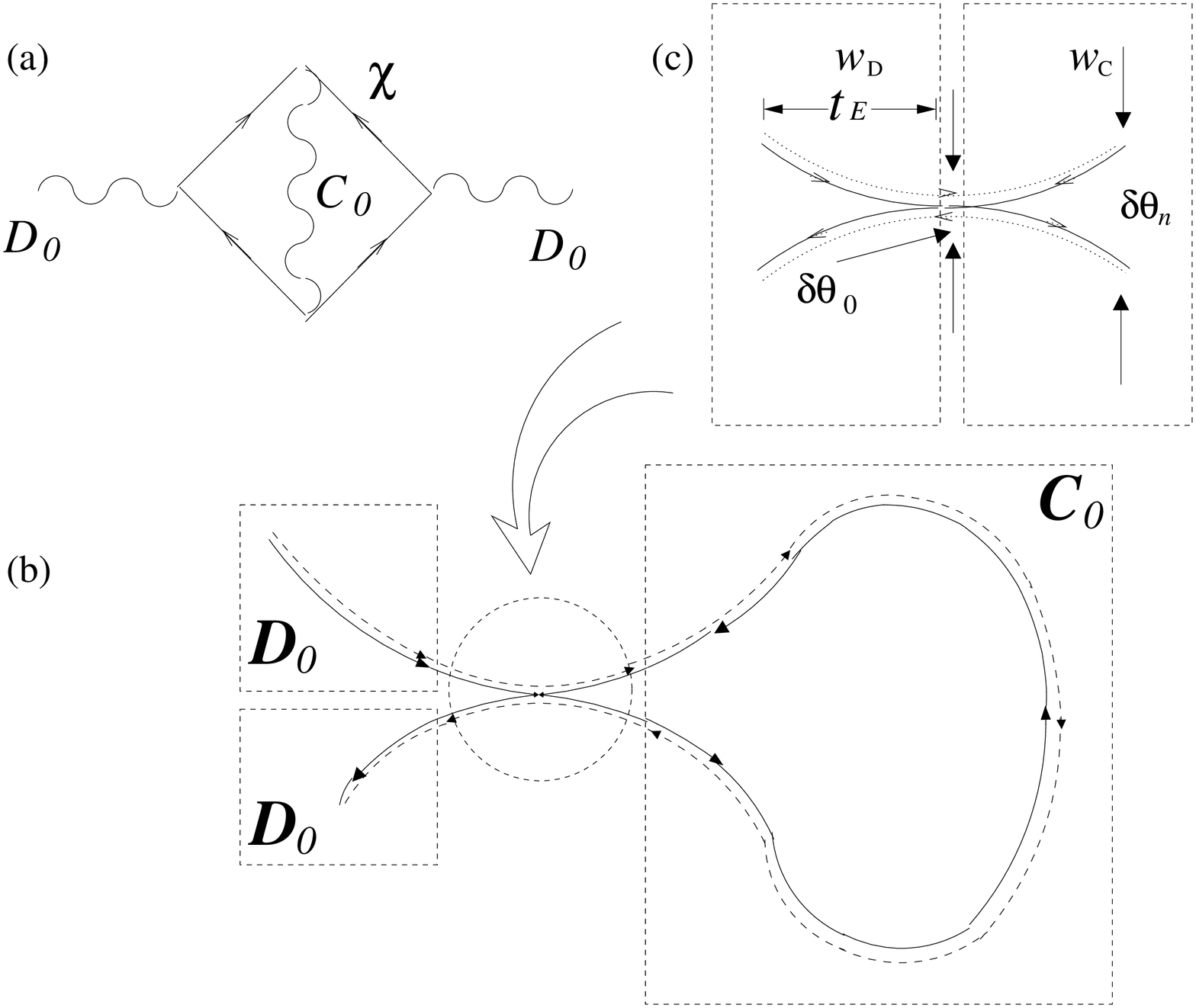}}
 \caption{The first quantum correction to the density--density
correlator: (a) one--loop weak localization diagram; (b)
trajectory in the momentum space; (c)  Hikami box along with
Lyapunov portions of the Cooperon and Diffusons. } \label{fig2}
\end{figure}

The physics of the weak--localization corrections is traditionally
discussed in the language of classical trajectories. The classical
motion of a particle in the random potential is characterized by a
rapid randomization of momenta and diffusion spreading of the
coordinate. It is thus customary to visualize a trajectory in the
coordinate space as a random motion between static impurities. It
is straightforward to develop a similar approach for the QKR. In
the kicked rotor problem the roles of coordinate and (angular)
momentum are interchanged. Indeed, for $K\gg 1$ the angular
coordinate  $\theta$ is rapidly randomized (over the interval
$[-\pi,\pi]$), while the angular momentum $l$ acquires a (quasi)
random change $\in[-K,K]$. The latter results in the diffusion in
the space of angular momentum (see below). We shall thus visualize
a ``trajectory'' as a sequence of values of the angular momentum
that kicked rotor ``visits'' upon successive kicks, Fig.
\ref{fig2}b.

For a quantitative description of the classical motion it is
convenient to monitor pairs of angle and angular momentum in
discrete moments of time. This maps the classical dynamics onto a
so-called standard map:
\begin{eqnarray}
l_{n+1}  &   =  &   l_n+K\sin\theta _n,\nonumber\\
\theta _{n+1}  &   =  &   \theta _n+ l_{n+1}\, . \label{StandardM}
\end{eqnarray}
Notice that $l_n$ stands for the angular momentum immediately
after $(n-1)$-th kick, and $\theta_n$ for the angle before $n$-th
kick. It is now obvious that the two successive points of the
trajectory: $l_n$ and $l_{n+1}$ differ by $K\sin\theta_n$. As a
first approximation one may treat $\theta_n$ as uniformly
distributed over $[0,2\pi]$ and thus
$\langle(l_{n+1}-l_n)^2\rangle=K^2\langle
\sin^2\theta\rangle=K^2/2$. As a result, $\langle
(l_n-l_0)^2\rangle =2D_{cl}n$ with $D_{cl}=K^2/4$. An account of
the residual correlations between successive $\theta_n$'s, leads
to a renormalization of $D_{cl}$ with the next term scaling as
$K^{3/2}$ etc. \cite{Rec81,LL}. For completeness of the
presentation we shall derive the full result for the classical
diffusion constant in Appendix \ref{Difffewkick}. For the current
qualitative discussion it is enough to appreciate that a
trajectory of the classical kicked rotor exhibits random hops in
the space of angular momentum, leading to:
\begin{equation}
\delta \langle l^2(t)\rangle \sim 2D_{cl}\, t\, .
\label{EnergyGrowthJoel}
\end{equation}
This result has a simple physical interpretation: the average
energy of the kicked rotor linearly increases with time,
reflecting a constant rate Joel's heating. It is exactly this
property of the classical kicked rotor that made it useful in the
accelerator physics \cite{Chirikov79,LL}.

In the quantum problem an {\em amplitude} to evolve from an
initial to a finite point in the angular momentum space is given
by the sum of {\em amplitudes} of all classically allowed
trajectories passing through these two points. Generically,
different trajectories come with random and uncorrelated phases
and thus do not produce a systematic interference contribution. An
exception to this rule comes from the trajectories having almost
(up to $\kbar$) exactly the same geometrical length and thus the
same phase. This situation takes place if a trajectory contains a
self--intersection point in the angular momentum space. Then
another trajectory may exist that is identical to the initial one
safe for the direction of  propagation along the loop, see Fig.
\ref{fig2} b. The fact that the backward propagation along the
loop is consistent with the equations of motion is guaranteed by
the time--reversal symmetry  of the Hamiltonian:
\begin{equation}
l\rightarrow l,\quad \theta\rightarrow -\theta, \quad t\rightarrow
-t, \quad H\rightarrow H\, . \label{TimSym}
\end{equation}
(compare it with the time-reversal symmetry in the random
potential problem: ${\bf r}\rightarrow {\bf r}\, ,{\bf
p}\rightarrow -{\bf p}\, , t\rightarrow -t$). It is thus easy to
see that we are interested in the loops that not only have
(almost) coinciding initial and finite momenta: $l_1\approx l_0$,
but also (almost) opposite initial and finite angles:
$\theta_1\approx -\theta_0$. The allowed uncertainty is limited by
the effective Planck constant:
$(l_1-l_0)(\theta_1+\theta_0)\lesssim \kbar$.

Since such two trajectories have (almost) the same phase, they
interfere constructively and thus lead to a systematic
(localization) quantum correction. The probability to complete the
loop in time $t$ is called Cooperon and  denoted as ${\cal
C}(l_0,\theta_0;l_1,\theta_1;t)$. With the above mentioned
conditions on the initial and finite points one may estimate it as
${\cal C}(t)\sim \sqrt{D_{cl}/t}$ (this must be multiplied by
$\kbar$ to take into account the phase area of the allowed
uncertainty). This estimate translates (basically by the double
integration over time) to the $-\kbar\sqrt{D_{cl}}\,t^{3/2}$
correction \cite{Basko,Altland93} to the classical law:
$\delta\langle l^2(t)\rangle=2D_{cl}t\, $. At $t\sim
t_L=D_{cl}/\kbar^2$ the correction exceeds the classical result
and the QKR crosses over to the strong localization regime.

The qualitative reasoning given above repeats identically the one
employed in the discussion of a particle in the field of the
''quantum'' white--noise random potential. However, the kicked
rotor dynamics possess a very important  distinction from that of
the white--noise potential problem. The latter is the process
without a memory. Indeed, two classical trajectories that
identically retrace each other up to a certain point may take
completely different roots (in particular counter--propagating
ones) after a {\em single} ``quantum'' scattering event. After
completing the loop in the opposite directions according to the
classical random walk dynamics, another {\em single} ``quantum''
scattering makes the two trajectories to be identical again. These
two (actually four, since there are two trajectories involved)
quantum scattering events constitute the, so--called, Hikami box
\cite{Hikami81}, denoted by ${\cal X}$ in Fig. \ref{fig2}a.

Contrary  to this scenario, the kicked rotor scattering events are
purely classical, namely the free rotation of the angle. Indeed
the trajectory is uniquely defined by the standard map,
Eq.~(\ref{StandardM}), sequence. If two trajectories coincide {\em
exactly} at some point $(\theta_n,l_n)$, they  continue to be
identical (determined by Eq.~(\ref{StandardM})) forever. This
seemingly precludes any possibility to develop the
weak--localization scenario, outlined above. The way out of this
apparent paradox is to recall that the loop may be completed not
exactly, but rather up to a small uncertainty: $\delta l \delta
\theta \lesssim \kbar$. This small initial difference is magnified
(more precisely, exponentially increases) upon successive kicking
(Lyapunov instability), leading eventually to the two
counter--propagating diffusive roots. The situation is rather
similar to the localization in the field of the {\em classical}
large scale random potential (so-called random Lorentz gas). The
latter was considered by Aliener and Larkin some time ago
\cite{AL96}. Due to complexity of the Lorentz gas dynamics, they
had to introduce quantum impurities (essentially a weak
white--noise component of the scattering potential) to treat the
problem analytically. The beauty of the QKR is in simplicity of
its classical dynamics, Eq.~(\ref{StandardM}), that allows to
treat the Lyapunov regime exactly without involving an artificial
quantum scattering.

To proceed in this direction, consider two trajectories that
initially  happen to be at a small distance from each other in the
phase space: $(\delta\theta,\delta l)$. Taking variation of
Eq.~(\ref{StandardM}), we find that the angle difference evolves
as
$\delta\theta_{n}=\delta\theta_{n-1}(1+K\cos\theta_{n-1})+\delta
l_{n-1}$\,. In the limit $K\gg 1$ it leads to
$\delta\theta_{n}=(\delta\theta+\delta l/K)\, \prod_{k=0}^{n-1}\,
(1+K\cos\theta_{k}) \simeq (\delta\theta+\delta l/K) e^{\lambda
n}$, where $\lambda$ is the Lyapunov exponent. For $K\gg 1$ one
finds: $\lambda=\langle\ln(K\cos\theta)\rangle = \ln(K/2)$\,
\cite{Chirikov79,LL} with the residual term $ \sim O(1/K)$. For
$\delta\theta\delta l\approx \kbar$, the optimal value of
$(\delta\theta+\delta l/K)$ is $\sqrt{\kbar/K}$. It thus takes
time $n=t_E=\lambda^{-1}\ln \sqrt{K/\kbar}$, cf.
Eq.~(\ref{ehrenfest}), to evolve from the initial angular
uncertainty $\delta\theta\approx \sqrt{\kbar/K}\ll 1$ up to
$\delta\theta_n\approx 1$, when the  diffusive motion takes over.
Once this deviation is reached, the usual diffusive spread of the
two trajectories takes place. The time--reversal invariance
dictates that aforementioned divergence of the two trajectories is
preceded by their convergence. The latter takes another $t_E$
kicks to be completed. The total duration of the one--way travel
through the Lyapunov region is thus $2t_E$. The entire
weak--localization loop construction requires two such travels
(each including convergence and divergence) through the Lyapunov
regime. As a result, the localization corrections are delayed by
(can not be developed in time less than) $4t_E$, see
Eq.~(\ref{resultApp}).

Technically there are two equivalent ways to incorporate the
Lyapunov region into the weak--localization calculations, see Fig.
\ref{fig2}c. One approach, adopted in Ref.~\onlinecite{AL96}, is
to redefine Hikami box to contain $4t_E$ scattering events (kicks)
instead of conventional four. Then Cooperon (and the Diffusons) is
just a conventional diffusive propagator in the momentum space. In
the present paper we find convenient to follow the traditional
treatment of Hikami box as consisting of the four kicks. These
four kicks are treated exactly by multiplying the corresponding
quantum evolution operators. It is the analytical expression for
Hikami box that dictates allowed deviations of $\delta
l\delta\theta$ product. The Lyapunov regimes are delegated to the
``legs'' of the Cooperon (and Diffusons). The latter now
understood as a propagator that includes both Lyapunov--like
divergence--convergence of the close trajectories along with the
normal diffusion once a macroscopic deviation between them is
reached. As we show below, the choice of the four kicks in the
Lyapunov regime, that coined to be Hikami box is immaterial. For
any such choice the quantum correction, linear in $\kbar$, is
delayed by  $4t_E$. It is important to mention that the time
interval $0  <  t  <  4t_E$ is protected from the higher order
loop corrections as well. To demonstrate this fact we performed
the two--loop weak--localization calculation and found the
corresponding contribution $\delta\langle l^2\rangle \sim
D_{cl}\theta(t-6t_E)(t-6t_E)^2/t_L$ to be delayed by $6t_E$.

The delay is not absolutely sharp, but  rather is slightly smeared
by  $\delta t_E \equiv \sqrt{\lambda_2 t_E/\lambda^2}\ll t_E$
number of kicks. The reason for this smearing is in fluctuations
of the exponent $\lambda$. Such fluctuations are due to the fact
that one follows the Lyapunov instability for a finite number of
kicks only. (Indeed, unlike classical problems, the minimal
deviation is limited by $\kbar$ and thus the  time to leave the
Lyapunov regime is finite and may fluctuate between the
trajectories). Following Ref.~\onlinecite{AL96} we characterize
fluctuations of the Lyapunov exponent by the other exponent,
$\lambda_2$ (for QKR with $K\gg 1$ one finds $\lambda_2\approx
.82$). In the case $t_E\gg 1$, cf. Eq.~(\ref{regime}), the effect
of smearing due to $\lambda_2$ is rather small.

\begin{figure}[h]
  \centerline{\epsfxsize=3in\epsfbox{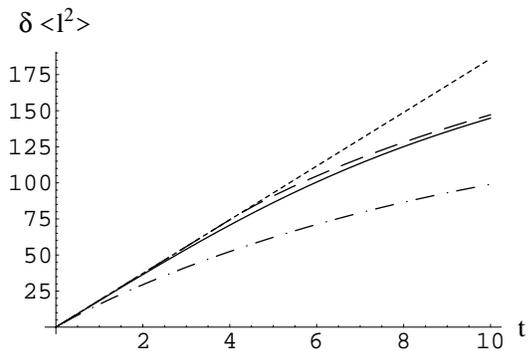}}
 \caption{The momentum dispersion for $K=6.1$ and  $\kbar =0.6$ --
full line; the classical limit ($\kbar \rightarrow 0$) -- dashed
line,  standard weak localization ($t_E=0$) -- dashed-dotted line;
the limit $\lambda_2\rightarrow 0$, Eq.~(\ref{resultApp}) --
long-dashed line.} \label{fig1}
\end{figure}

The predicted  time--dependent momentum dispersion graph is
plotted on Fig.~\ref{fig1}.  The following sections serve to
quantify the qualitative semiclassical picture outlined above.

\section{Classical limit: Diffuson and Cooperon}
\label{classicalDC}

We proceed now to develop the qualitative considerations outlined
above into an accurate theory of the  QKR. The essential starting
point is the classical diffusion in the angular momentum direction
of the classical phase space: $[-\pi\, ,+\pi]\times [-\infty\, ,
+\infty ]$\,. We'll clarify first how to find the  classical
diffusion from the exact quantum correlator. One starts from
introducing the exact one period quantum evolution operator as:
\begin{eqnarray}
  &    &   \hat U\equiv {\hat V}\, {\hat J}\, , \label{U}\\
  &    &   {\hat V}=\exp{(i{\hat l}^2/2\kbar)},\, \quad {\hat
J}=\exp{(iK\cos {\hat \theta}/\kbar)} \, . \nonumber
\end{eqnarray}
All physical quantities may be expressed  in terms of the matrix
elements of $\hat U^n$, where $n$ stays for the number of kicks
(time). We are particularly interested in the four--point
``density--density'' correlator, defined as:
\begin{eqnarray}
   &   &   {\cal D}(l_+,l_-; l_+',l_-';n,n') \nonumber\\
    &   \equiv  &   \langle l_+| {\hat U}^{n}
 e^{ \frac{i\hat l ^2}{2\kbar}}|l_+' \rangle
\langle l_-|{\hat U}^{n'} e^{\frac{i\hat l ^2}{2\kbar}}|l_-'
\rangle ^*\, ,
\label{QDiff}
\end{eqnarray}
where $|l_{\pm}\rangle$ denote basis of discrete momentum
eigenstates of $K=0$ quantum Hamiltonian. Note that  the forward
and backward trajectories in this expression are, in general,
different. However, it is natural to expect after some transient
processes, the correlator will be dominated by the case of $n=n'$.
This is easy to see  by Fourier transforming the correlator with
respect to $n\, , n'$, passing to the frequency $(\omega_+\, ,
\omega_- )$ representation:
\begin{eqnarray}
  &    &   {\cal D}(l_+,l_-; l_+',l_-';\omega_+,\omega_-)
\label{QuantumDFourier}\\
  &  =  &   \sum_{n,n'=0}\, e^{i(\omega_+ n-\omega_- n')}\, {\cal
D}(l_+,l_-; l_+',l_-';n,n') \,  \nonumber
\end{eqnarray}
and subsequent averaging over $(\omega_++\omega_-)/2$. Let us
denote such correlator thereby obtained as ${\cal D}(l_+,l_-;
l_+',l_-';\omega) $, where $\omega = \omega_+-\omega_-$. From
Eq.~(\ref{QDiff}) one may check that it satisfies
\begin{eqnarray}
  &    &  {\cal D}(l_+,l_-; l_+',l_-';\omega)=
e^{\frac{i\left(l_+^2-l_-^2\right) }{2\kbar}} \, \delta_{l_+,\,
l_+'}\delta_{l_-,\, l_-'}
                                               \label{DDDefSig}\\
  &    &   + e^{i\omega} \sum_{l_+'',\, l_-''} \langle l_+|{\hat
U}|l_+'' \rangle
\langle l_-|{\hat U}|l_-'' \rangle ^* \, {\cal D}(l_+'',l_-'';
l_+',l_-';\omega)\, . \nonumber
\end{eqnarray}
The matrix elements $\langle l_+|{\hat U}|l_+'' \rangle$ and
$\langle l_-|{\hat U}|l_-'' \rangle ^*$ may be explicitly written
as
\begin{eqnarray}
  &    &   \langle l_+|{\hat U}|l_+'' \rangle
\langle l_-|{\hat U}|l_-'' \rangle ^* \nonumber\\
  &  =  &
\int\!\!\!\int\frac{d\theta_+}{2\pi}\frac{d\theta_-}{2\pi} \exp
\left[\frac{il_+^2}{2\kbar}+\frac{iK\cos\theta_+}{\kbar}+\frac{i\theta_+}{\kbar}
\left(l_+-l_+'' \right) \right] \nonumber\\
  &    &  \times \exp
\left[-\frac{il_-^2}{2\kbar}-\frac{iK\cos\theta_-}{\kbar}-\frac{i\theta_-}{\kbar
} \left(l_--l_-'' \right) \right]\, . \label{selfenergyPerron}
\end{eqnarray}
For what follows, it is convenient to introduce the Wigner
transform representation as:
\begin{eqnarray}
  &    &  {\cal D} \left(l,\theta;l',\theta';\omega \right)\nonumber\\
  &  \equiv  &   \sum_{l_+-l_-} \sum_{l_+'-l_-'}\exp \left(-
\frac{i}{\kbar}\left[(l_+-l_-)\theta -(l_+'-l_-')\theta'\right]
\right) \nonumber\\
  &    &  \times {\cal D}(l_+,l_-; l_+',l_-';\omega)\, ,
\label{DWig}
\end{eqnarray}
where we define $l\equiv (l_++l_-)/2$ and $l'\equiv
(l_+'+l_-')/2$.

\subsection{Frobenius-Perron-Ruelle equation: classical kicked rotor}
\label{CLKR}

Let us consider a solution of Eq.~(\ref{DDDefSig}). As we will see
below, the integral over $\theta_\pm$ in
Eq.~(\ref{selfenergyPerron}) are dominated by $|\theta_+-\theta_-|
\sim \kbar/K \ll 1$. This then allows for a perturbative expansion
of $|\theta_+-\theta_-|$ in the exponent of
Eq.~(\ref{selfenergyPerron}). In this subsection we show that the
leading term in such expansion leads to the classical equation of
motion (standard map), Eq.~(\ref{StandardM}) (so-called
``semiclassical approximation''). The semiclassical solution
thereby obtained is denoted as ${\cal D}_0$. It should be
emphasized that the classical diffusive propagator can not be
recovered at this stage. To achieve this goal, the further
approximation must be used, which will be clarified in
Sec.~\ref{DiffApprox}.

The semiclassical treatment employs the following approximation
for  $\langle l_+|{\hat U}|l_+'' \rangle \langle l_-|{\hat
U}|l_-'' \rangle ^* $ matrix elements, cf.
Eq.~(\ref{selfenergyPerron}),
\begin{widetext}
\begin{eqnarray}
  &    &   \langle l_+|{\hat U}|l_+'' \rangle \langle l_-|{\hat
U}|l_-'' \rangle ^* \approx \int\!\!\!
\int\frac{d\theta_+}{2\pi}\frac{d\theta_-}{2\pi} \exp
\left[\frac{i(l_++l_-)(l_+-l_-)}{2\kbar}-\frac{iK}{\kbar}\sin\frac{\theta_++\theta_-}{2}
\left(\theta_+-\theta_-\right)\right] \nonumber\\
  &  \times  &
\exp\left\{\frac{i}{\kbar}\left(\theta_+-\theta_-\right)\left(\frac{l_++l_-}{2}-
\frac{l_+''+l_-''}{2}\right)+
\frac{i}{\kbar}\frac{\theta_++\theta_-}{2}\left[\left(l_+-l_-\right)-\left(l_+''
-l_-''\right)\right] \right\} \, . \label{selfenergyPerronsemi}
\end{eqnarray}
\end{widetext}
\noindent Let us insert it into Eq.~(\ref{DDDefSig}) and perform
the Wigner transform. We also define $\theta \equiv
(\theta_++\theta_-)/2$ and $\theta' \equiv
(\theta_+'+\theta_-')/2$ to simplify the notations. Then with
$(\theta_+-\theta_-)/\kbar$, $(l_+-l_-)/\kbar$,
$(l_+'-l_-')/\kbar$ and $(l_+''-l_-'')/\kbar$ integrated out, we
obtain:
\begin{eqnarray}
{\cal D}_0(l,\theta;l',\theta';\omega)   &  =  &   2\pi\kbar\,
\delta(l-l')\, \delta(\theta -\theta '-l) \nonumber\\
  &    &   + e^{i\omega} \overrightarrow{P} {\cal
D}_0(l,\theta;l',\theta';\omega) \, , \label{DysonCorrelator}
\end{eqnarray}
where $\overrightarrow{P}$ is the Frobenius-Perron-Ruelle (FPR)
operator, acting on the nearest two arguments from the left
according to:
\begin{eqnarray}
  &    &   \overrightarrow{P}\, f \left(l,\theta;l',\theta'
\right)
\nonumber\\
  &  \equiv  &   \int\!\!\! \int dl_1\, d\theta_1 \,
\delta\left(l-l_1-K\sin\theta_1 \right)
\delta\left(\theta-\theta_1-l\right) \nonumber\\
  &    &   \times f \left(l_1,\theta_1;l',\theta' \right)\, ,
\label{PFDef}
\end{eqnarray}
where  $f\left(l,\theta;l',\theta' \right)$ is an arbitrary
function. The kernel above implies that the correlator thereby
obtained describes the deterministic motion of {\it classical}
kicked rotor, i.e., standard mapping. Note, that in the time
representation, ${\cal D}_0$ is normalized, namely
$(2\pi\kbar)^{-1 }\int\!\!\int dld\theta {\cal
D}_0(l,\theta;l',\theta';n)=1$.

\subsection{Diffusion approximation}
\label{DiffApprox}

To recover the classical diffusion the procedure of deriving FPR
equation must be appropriately regularized
\cite{Rec81,KFA00,Zirnbauer99}. Indeed, in the presence of noises
one is able to do so and find the proper diffusion constant. We
shall not follow this procedure here, but rather refer a reader to
Refs.~\onlinecite{Rec81,KFA00}.

We shall show  below, however, that a regularization is consistent
with the generalization of Altland's \cite{Altland93} diagrammatic
method. The latter starts from exact quantum density-density
correlator, Eq.~(\ref{QDiff}), and leads to the classical
diffusion with the correct diffusion constant. The basic idea  is
that  for sufficiently large $K$, there must be some time scale,
say $\tau_c$, beyond which any classical trajectory loses memory
about its initial conditions. Technically, complementing the
semiclassical approximation: $|\theta_{n_+}-\theta_{n_-}|\ll 1\, ,
n=1,2,\cdots$, it is further required that $l_+=l_-\equiv l$ and
$l_+'=l_-'\equiv l'$ at two ends, as well as $l_+''=l_-''\equiv
l''$ for the intermediate variables at multiple times of $\tau_c$.

\begin{figure}[h]
  \centerline{\epsfxsize=3in\epsfbox{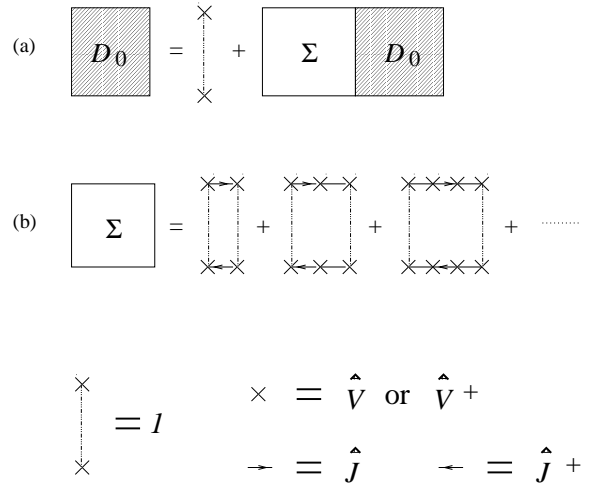}}
\caption{ Schematic representation for the diffusion approximation
of classical density-density correlator -- the solution of
Frobenius-Perron-Ruelle equation (a), with the self-energy (b).
The structure of the self-energy is such that the  beginning and
ending (after $\tau_c$ kicks) pairs of  the angular momenta  are
the same (denoted by vertical dot-dashed line), while the other
pairs not. } \label{fig3}
\end{figure}

For $\tau_c=1$\, the angular memory  is lost after every kick (cf.
the first diagram on the right hand side of Fig.~\ref{fig3} b). In
this  approximation  Eq.~(\ref{DDDefSig}) is reduced to
\begin{eqnarray}
  &    &  {\cal D}_0(l, l';\omega)\nonumber\\
  &  =  &   \delta_{l,\, l'} + e^{i\omega} \sum_{l''} \langle
l|{\hat U}|l'' \rangle
\langle l|{\hat U}|l'' \rangle ^* \, {\cal D}_0(l'',l';\omega) \,
, \label{DiffPropoOneKick}
\end{eqnarray}
where the matrix element $\langle l|{\hat U}|l'' \rangle \langle
l|{\hat U}|l'' \rangle ^*$ is explicitly written as
\begin{eqnarray}
  &    &   \langle l|{\hat U}|l'' \rangle \langle l|{\hat U}|l''
\rangle ^* = \langle l|{\hat J}|l'' \rangle \langle
l|{\hat J}|l'' \rangle ^* \nonumber\\
  &  =  &   \int\!\!\!
\int\frac{d\theta_+}{2\pi}\frac{d\theta_-}{2\pi} \exp
\left[\frac{iK}{\kbar}\left(\cos\theta_+-\cos\theta_-\right)\right]
\nonumber\\
  &    &  \times \exp
\left[\frac{i\left(\theta_+-\theta_-\right)\left(l-l''\right)}{\kbar}\right]
\nonumber\\
  &  =  &   \int\!\!\!
\int\frac{d\theta_+}{2\pi}\frac{d\theta_-}{2\pi} \exp
\left[-\frac{2iK}{\kbar}\sin\frac{\theta_++\theta_-}{2}\sin\frac{\theta_+-\theta
_-}{2}\right]
\nonumber\\
  &    &  \times \exp
\left[\frac{i\left(\theta_+-\theta_-\right)\left(l-l''\right)}{\kbar}\right]
\, . \label{selfenergydiff}
\end{eqnarray}
From this we see (as mentioned above) that the integral is
dominated by $|\theta_+-\theta _-| \sim \kbar/K $. Making the
change of variables: $(\theta_+\, ,\theta_-)\rightarrow
[(\theta_++\theta_-)/2, \theta_+-\theta_- \equiv \kbar\varphi]$
and integrating out $(\theta_++\theta_-)/2$, we simplify
Eq.~(\ref{selfenergydiff}) as\, \cite{Altland93}
\begin{equation}
\langle l|{\hat U}|l'' \rangle \langle l|{\hat U}|l'' \rangle ^*
\approx \int \frac{d\varphi}{2\pi}
J_0\left[\frac{2K}{\kbar}\sin\frac{\kbar\varphi}{2}\right]
e^{i\varphi\left(l-l''\right)} \, . \label{selfenergyAltland}
\end{equation}
Here $J_n(x)$ is the Bessel function of order $n$. Turning to the
Fourier representation: ${\cal D}_0(\varphi;\omega)\equiv
\sum_{l-l'} \, e^{-i\varphi (l-l')}\, {\cal D}_0(l-l';\omega)$,
Eq.~(\ref{DiffPropoOneKick}) gives the classical propagator as
\begin{equation}
{\cal
D}_0(\varphi;\omega)=\frac{1}{1-e^{i\omega}J_0\left[\frac{2K}{\kbar}\sin
\frac{\kbar\varphi}{2}\right]} \, . \label{DiffClassict1}
\end{equation}
In the limit $\omega\ll 1\, , K\varphi\ll 1$, it is reduced to the
usual  diffuson (Fig.~\ref{fig3}):
\begin{equation}
{\cal D}_0(\varphi;\omega)\approx \frac{1}{-i\omega  +D_{cl}\,
\varphi^2} \label{DiffAltland}
\end{equation}
with the diffusion constant $D_{cl}(K)=K^2/4$.

In Ref.~\onlinecite{Rec81}, it was shown that the higher order
correlations, namely $\tau_c   >   1$ [e.g., the second, third,
etc. diagrams on the right hand side of Fig.~\ref{fig3}b] lead to
the modification of the diffusion constant according to
\begin{equation}
D_{cl}(K)=\frac{K^2}{4}\left[1-2J_2(K)-2J_1^2(K)+2J_2^2(K)+\cdots\right]
\, . \label{Dcl}
\end{equation}
We should emphasize that although the original derivation is based
on pure classical considerations, it is fully compatible with the
general formalism developed in this paper. For the clarification,
we reproduce Eq.~(\ref{Dcl}) from the exact quantum
density--density correlator, Eq.~(\ref{DDDefSig}), in
Appendix~\ref{Difffewkick}.

One may wonder whether the dynamics of a classical system may be
indeed described by the diffusion in the long time limit. The
answer is known to depend on the initial conditions. In
particular, it is well--known that for any $K$, there exist stable
islands in the phase space, where a trajectory exhibits
quasi-periodic motion \cite{Chirikov79,LL}.  It has been
estimated, however, that the total area of these islands on the
phase space is exponentially small in the limit $K\gg 1$
\cite{Chirikov79,Izrailev90}. Our approximation, being an
expansion in powers of $1/K$, is bound to lose information about
these islands. It has also been known that for some parametric
regions of $K$, there are some peculiar islands (so-called
``accelerator mode'' \cite{Chirikov79,LL}), starting from or near
which a trajectory will be boosted, faster than the normal
diffusion \cite{Zaslavsky97}. ( Similar phenomenon exist also in a
generic model \cite{LL,Raizen00}. )  Throughout the paper we will
stay away from these parametric regions, and focus on the large
$K$ case with the assumption that the initial conditions do not
fall neither into the stable islands nor to the accelerator modes.

\subsection{Classical Cooperon}
\label{coop}

Starting from the exact quantum density--density correlator, one
may see that the classical diffuson is essentially related to the
limiting case: $|\theta_{1+}-\theta_{1-}|\sim
|\theta_{2+}-\theta_{2-}|\cdots \sim |\theta_{n+}-\theta_{n-}| \ll
1$. In this part, we turn to the discussion of another important
limiting case, i.e., $|\theta_{1+}+\theta_{n-}|\sim
|\theta_{2+}+\theta_{(n-1)-}|\cdots \sim
|\theta_{n+}+\theta_{1-}|\ll 1$ (cf. Fig.~\ref{Cooperonfig} for
notations).

\subsubsection{Diffusion approximation of the solution of FPR equation}
\label{GeneralCooperon}

\begin{figure}[h]
  \centerline{\epsfxsize=3in\epsfbox{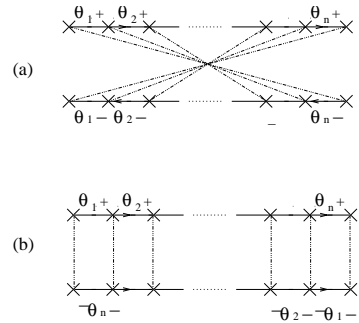}}
 \caption{ The diffusive Cooperon approximation for  the
classical density-density correlator. Typical diagrammatic
representation for the classical Cooperon
($|\theta_{1+}+\theta_{n-}|\, ,|\theta_{2+}+\theta_{(n-1)-}|\,
,\cdots\ll 1$) (a). Rotating the bottom line of (a) retrieves
classical diffuson with $\theta_{k-}$ replaced by $-\theta_{k-}$
($k=1,2,\cdots, n$) (b). } \label{Cooperonfig}
\end{figure}

To find the density-density correlator in this limit let us pass
to Wigner representation (notice the crucial difference with
Eq.~(\ref{DWig})):
\begin{eqnarray}
  &    &  {\cal C} \left(l,\theta;l',\theta';\omega \right)\nonumber\\
  &  \equiv  &   \sum_{l_+-l_-'} \sum_{l_+'-l_-}\exp \left(-
\frac{i}{\kbar}\left[(l_+-l_-')\theta -(l_+'-l_-)\theta'\right]
\right) \nonumber\\
  &    &  \times {\cal C}(l_+,l_-; l_+',l_-';\omega)\, ,
\label{CWig}
\end{eqnarray}
where we define $l\equiv (l_++l_-')/2$ and $l'\equiv
(l_+'+l_-)/2$. The key observation is that the diffuson diagram
may be retrieved with the bottom (advanced Green function) line
[Fig.~\ref{Cooperonfig} (a)] rotated. This is due to the fact that
such a procedure simply leads to the time-reversal of the series:
$\theta_{1-}\rightarrow \theta_{2-}\rightarrow \cdots
\rightarrow\theta_{n-}$ such that it becomes
$-\theta_{n-}\rightarrow -\theta_{{n-1}-}\rightarrow \cdots
\rightarrow-\theta_{1-}$\,. For this reason, we call it Cooperon
as introduced in Sec.~\ref{GenCon}. Since
$|\theta_{1+}+\theta_{n-}|\ll 1$, we may proceed along the lines
of derivation of Eq.~(\ref{DysonCorrelator}) from
Eqs.~(\ref{DDDefSig}) and (\ref{DWig}), and arrive at
\begin{eqnarray}
{\cal C}_0(l,\theta;l',\theta';\omega)  &  =  &   2\pi\kbar\,
\delta(l-l')\, \delta(\theta -\theta '-l) \nonumber\\
  &    &   + e^{i\omega} \overrightarrow{P} \, {\cal
C}_0(l,\theta;l',\theta';\omega) \, . \label{CooperonCorrelator}
\end{eqnarray}
That is, the classical Cooperon is also a solution of FPR
equation.

The only difference with the Diffuson is that instead of having
poles at $\kbar\varphi\equiv \theta_+-\theta_-$, the Cooperon has
poles at $\kbar\varphi\equiv \theta_++\theta_-$ (i.e.,
$\kbar\varphi \equiv \theta_{1+}+\theta_{n-} =
\theta_{2+}+\theta_{(n-1)-} = \cdots = \theta_{n+}+\theta_{1-}$).
One concludes, thus, that the diffusive form,
Eq.~(\ref{DiffAltland}), holds in the limit $\omega\ll 1\, ,
K\varphi\ll 1$ (cf. Fig.~\ref{Cooperonfig}) for the Cooperon as
well. In particular, in the case of $|\theta+\theta'|\sim 1$\,,
the averaging over $(\theta+\theta')/2$ may be performed.
Furthermore, if $|l-l'|\lesssim K$\,, then ${\cal C}_0$ does not
depend on the angular momenta. In other words,
\begin{eqnarray}
  &    &   {\cal C}_0\left(l,\theta;l',-\theta';\omega\right)  =
\left\langle{\cal C}_0\left(l,\theta;l',-\theta';\omega\right)
\right\rangle_{(\theta+\theta')/2} \nonumber\\
  &    &   = \int \frac{d\varphi}{2\pi}\, \frac{\kbar}{
-i\omega +D_{cl}\, \varphi ^2} \equiv \left\langle {\cal C}_0
\left(\omega\right)\right\rangle \, , \label{coopBC}
\end{eqnarray}
where $|l-l'|\lesssim K,\,$ and  $|\theta-\theta'|\sim 1$\,.

\subsubsection{Treatment of the Lyapunov instability regime}
\label{RenormaDiffuCooperon}

The above general solution for the  classical Cooperon, ${\cal
C}_0(l,\theta;l',-\theta'; n)$\,, characterizes the probability
for a trajectory, initiating from $(l,\theta)$, to end at
$(l',-\theta')$. From now on we focus on a special case, where
$\delta l_0\equiv l-l'\, ;\delta \theta_0\equiv \theta-\theta' $
are such small that $|\delta l_0|\ll K,\, |\delta \theta_0|\ll
1$\,. In this part we show that it differs from $\langle {\cal
C}_0 (\omega)\rangle$ by a renormalization factor.

Without loss of generality, we assume that ${\cal
C}_0(l,\theta;l',-\theta'; n)$ evolves from some initial
distribution $f\left( l,\theta;l',-\theta' \right)$, bearing the
symmetry of $ f\left( l,\theta;l',-\theta' \right)=f\left(
l',\theta';l,-\theta \right)$\,. Then the formal solution of the
FPR equation is given by
\begin{equation}
{\cal C}_0\left(l,\theta;l',-\theta';n\right)
=\overrightarrow{P}\, ^n\,f\left( l,\theta;l',-\theta' \right) \,
. \label{Cooperon}
\end{equation}

To simplify the notations, we rewrite the action of FPR operator
on ${\cal C}_0$ as:
\begin{equation}
\overrightarrow{P}\, {\cal C}_0 \left(l,\theta;\, l',\theta'
\right)= {\cal C}_0 \left({\cal S}^{-1} \left[l,\theta\right] ;\,
l',\theta' \right) \, , \label{leftP}
\end{equation}
where ${\cal S}$ and ${\cal S}^{-1}$ are defined as
\begin{eqnarray}
{\cal S}\left[l\,, \theta \right]   &  \equiv  &
\left(l+K\sin(\theta+l), \,
\theta +l \right) \, , \nonumber\\
 {\cal S}^{-1}\left[l\,, \theta
\right]   &  \equiv  &   \left(l-K\sin(\theta -l), \, \theta -l
\right) \, , \label{inverseStanM}
\end{eqnarray}
respectively. Moreover, we introduce the time-reversal of FPR
operator, $\overleftarrow{P}_T$, as
\begin{eqnarray}
f \left(l,\theta;l',\theta'\right)\,\overleftarrow{P}_T \,
\equiv\int\!\!\! \int dl_1\, d\theta_1 \, f
\left(l,\theta;l_1,\theta_1\right) \nonumber\\
\times \delta\left(l_1-l'-K\sin\theta_1 \right)
\delta\left(\theta_1-\theta'-l\right) \, . \label{PFTDef}
\end{eqnarray}

Owing to the symmetry of $f$, mentioned above, one may introduce
the following identity:
\begin{equation}
\overrightarrow{P}\, f \left(l,\theta;\, l',\theta' \right)= f
\left(l,\theta;\, l',\theta' \right) \, \overleftarrow{P}_T\, ;
\label{leftrightPrela}
\end{equation}
its proof is given in  Appendix~\ref{relation}. Applying
repeatedly  this relation to the formal solution,
Eq.~(\ref{Cooperon}), we obtain:
\begin{eqnarray}
  &    &   {\cal C}_0\left(l,\theta;l',-\theta';n\right) \nonumber\\
  &  =  &   \overrightarrow{P}\,^{n-2n'}\, \left\{
\overrightarrow{P}\,^{n'}\, f\left( l,\theta;l',-\theta'
\right) \overleftarrow{P}_T ^{n'}\, \right\} \nonumber\\
  &  =  &   \overrightarrow{P}\,^{n-2n'}\, f\left( {\cal
S}^{-n'}[l,\theta] \, ; {\cal S}^{n'} [l',-\theta'] \right)
\label{CooperonEvolution}
\end{eqnarray}
for an arbitrary integer number $n'$ such that $2n'\leq n$\,.
Consider two nearby trajectories described by
$\left(l_1,\theta_1\right)$ and $\left(l_1',\theta_1'\right)$\,,
respectively. Their motion is induced by ${\cal S}^{-n}[l,\theta]$
and ${\cal S}^{n} [l',-\theta']$ following
\begin{eqnarray}
\left(l_1,\theta_1\right)  &  \equiv  &  {\cal S}^{-n} \,
[l,\theta] \, ,
\nonumber\\
\left(l_1',-\theta_1'\right)  &  \equiv   &   {\cal S}^{n}\,
[l',-\theta'] \, . \label{Snc}
\end{eqnarray}
Associated with the exponential separation of these two nearby
trajectories, the time, say $n_{\rm c}$, is defined such that
\begin{equation}
|l_{n_c}-l_{n_c}'|\approx K \, , \quad
|\theta_{n_c}-\theta_{n_c}'|\approx 1  \, .
\label{ncDefinition}
\end{equation}
At such a moment the separation reaches some macroscopic size.
After this time the separation experiences the usual diffusion in
the angular momentum space, while the angle difference is
uniformly distributed.  We substitute then $n'=n_{\rm c}$ into
Eq.~(\ref{CooperonEvolution}), and arrive at
\begin{equation}
{\cal C}_0\left(l,\theta;l',-\theta';n\right)= \theta
\left(n-2n_{\rm c}\right) \left\langle {\cal C}_0\left(n-2n_{\rm
c}\right) \right\rangle \, \label{CoopFactorization}
\end{equation}
where $\langle {\cal C}_0 (n) \rangle$ is the Fourier transform of
$\langle {\cal C}_0(\omega) \rangle$. Consequently, the Fourier
transform of Eq.~(\ref{CooperonEvolution}) with respect to $n$ is
reduced to
\begin{equation}
{\cal C}_0(l,\theta;l',-\theta';\omega) =e^{2i\omega n_{\rm c}} \,
\int \frac{d\varphi}{2\pi}\, \frac{\kbar}{-i\omega +D_{cl}\,
\varphi ^2} \, . \label{cooWZ0}
\end{equation}

In general, $n_{\rm c}$ is determined by the initial condition,
namely the center of mass: $ [(l+l')/2,(\theta+\theta')/2 ]$ and
the initial deviation: $\left(\delta l,\delta\theta\right)$\,. One
can perform the change of variables with respect to general
separation $\left(\delta l,\delta\theta\right)$ according to
$z\equiv\ln|\delta\theta|,\, \alpha\equiv \delta l/\delta
\theta$\,. In what follows $z$ and $\alpha$ are identified as slow
and fast variables, correspondingly.  If the initial deviation is
small enough such that the typical $n_{\rm c}\gg 1$, the
fluctuations of $n_{\rm c}$ at this time scale are small. As a
result, after averaging over initial $\alpha$, as well as the
center of mass, one may cast $\exp[2i\omega n_{\rm c}]$ into the
renormalization factor (with the logarithmic accuracy) as (see
Appendix~\ref{PropagatorLyapunov} for details)
\begin{eqnarray}
 {\cal W}_{\rm C}(2\omega)   &   \equiv   &   \left\langle \exp \left[2i\omega n_{\rm
c}\right]\right\rangle  \nonumber\\
   &   =   &    \exp \left(2i\omega t^{\rm
C}-\frac{2\omega ^2\lambda _2 t^{\rm C}}{\lambda ^2}\right)
\label{Gamma}
\end{eqnarray}
with
\begin{equation}
t^{\rm C}\equiv
\frac{1}{\lambda}\left|\ln\frac{1}{\sqrt{\delta\theta^2+\delta
l^2/K^2}}\right| \, . \label{EhrTimeS}
\end{equation}
in the limit $\omega\lambda _2/\lambda ^2\ll 1$\,. Consequently,
Eq.~(\ref{cooWZ0}) is reduced to
\begin{equation}
{\cal C}_0(l,\theta;l',-\theta';\omega) =\kbar \, {\cal W}_{\rm
C}(2\omega)\int \frac{d\varphi}{2\pi}\, \frac{1}{ -i\omega
+D_{cl}\, \varphi ^2} \, . \label{CoopRenor}
\end{equation}

\section{Weak dynamical localization in kicked rotor: One-loop Correction}
\label{WDLKR}

\begin{figure}[h]
  \centerline{\epsfxsize=3in\epsfbox{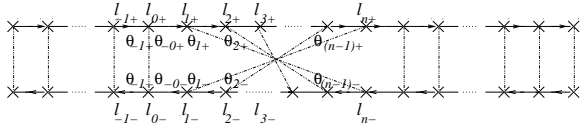}}
\caption{ Sketch of a general diagram, leading to one-loop
approximation. } \label{oneloopfig}
\end{figure}

As explained above the weak dynamical localization involves
couplings between the Diffusons and the Cooperon
[Fig.~\ref{oneloopfig}]. Therefore one needs a technique to treat
two different kinds of the Wigner transforms introduced for
Diffusons, Eq.~(\ref{DWig}), and for Cooperons, Eq. (\ref{CWig}),
in a unified way. To develop such a technique is the central task
of this section. We  show then that the constructive interference
between two counter--propagating trajectories leads to the usual
one--loop quantum correction, which is a precursor of the
dynamical localization. In particular, the one--loop correction to
the diffusion constant will be calculated.

\subsection{Exact interaction vertex}
\label{OLC}

We will show in Appendix~\ref{wardiden} that
the one-loop correction to the density-density correlator
reads as
\begin{equation}
\delta {\hat {\cal D}}= {\hat {\cal D}}\, \left(e^{i\omega}\,
{\hat {\cal P}}_T-1 \right) \, {\hat {\bf {\rm C}}}\,
\left(e^{i\omega}\, {\hat {\cal P}}-1 \right) \, {\hat {\cal D}}
\label{oneloopexact}
\end{equation}
with
\begin{equation}
{\hat {\bf {\rm C}}}\equiv e^{i\omega}\, {\hat {\cal P}}_J \,
{\hat {\cal C}}_0 \, e^{i\omega}\, {\hat {\cal P}}_J \label{CC}
\end{equation}
in the exact quantum operator representation. Here ${\hat {\cal
P}}\equiv \, {\hat {\cal P}}_V\, {\hat {\cal P}}_J$ and ${\hat
{\cal P}}_T \equiv \, {\hat {\cal P}}_J\, {\hat {\cal P}}_V$. The
matrix elements of ${\hat {\cal P}}_V$ and ${\hat {\cal P}}_J$ are
explicitly written as
\begin{eqnarray}
\langle l_+\, ,l_-|{\hat {\cal P}}_V|l_+'\, ,l_-'\rangle
  &  =  &   \exp \left[\frac{i\left(l_+^2-l_-^2\right)}{2\kbar}
\right] \,
\delta_{l_+, \,l_+'} \, \delta_{l_-, \,l_-'}\, ,\nonumber\\
\langle \theta_+\, ,\theta_-|{\hat {\cal P}}_J|\theta_+'\,
,\theta_-'\rangle   &  =  &   \exp
\left[\frac{iK\left(\cos\theta_+-\cos\theta_-\right)}{\kbar}
\right]\nonumber\\
  &    &  \times\delta\left(\theta_+-\theta_+' \right)\,
\delta\left(\theta_--\theta_-' \right) \label{VJmatrix}
\end{eqnarray}
in the representation of the angular momentum and the angle,
correspondingly.  Since we are ultimately interested in the long
time effects and therefore in the low frequencies the
$e^{i\omega}$ factors in Eqs.~(\ref{oneloopexact}) and (\ref{CC})
may be safely omitted from now on. Thus, we shall not write them
hereafter.

\subsubsection{Minimal wave packet}
\label{OLC}

In order to calculate
$\delta {\hat {\cal D}}$ explicitly, we consider a general
quantity, say ${\hat {\cal I}}_q\, \equiv\, {\hat {\cal A}} {\hat
{\bf {\rm C}}} {\hat {\cal B}}$ (understanding that ${\hat {\cal
A}}={\hat {\cal D}}\, \left( {\hat {\cal P}}_T-1 \right),\, {\hat
{\cal B}} = \left( {\hat {\cal P}}-1 \right) \, {\hat {\cal D}}$
for $\delta {\hat {\cal D}}$), and write it explicitly as (for
simplicity we omit $\omega$ argument)
\begin{eqnarray}
  &    &   {\cal I}_{q}
\left(l_+,l_-;l_+',l_-'\right)=\sum_{l_1,\, l'_1}\sum_{l_2,\,
l'_2}\sum_{l_3,\, l'_3}\sum_{l_4,\, l'_4}
\label{Iq} \\
  &  \times  &   {\cal A} \left(l_+,l_-;l_1,l_1'\right) \, {\cal
X}\, {\cal C}_0 \left(l_{2} ,l_{3} ;l_{2} ',l_{3} '\right)\, {\cal
B} \left(l_4,l_4';l_+',l_-'\right) \nonumber
\end{eqnarray}
in the angular momentum representation, where ${\cal X}$ is
\begin{eqnarray}
{\cal X}   &  =  &   \left\langle l_1\left|e^{i\frac{K}{\kbar}\cos
{\hat \theta}}\right|l_2\right\rangle \left\langle
l_1'\left|e^{i\frac{K}{\kbar}\cos {\hat
\theta}}\right|l_2'\right\rangle ^* \nonumber\\
  &    &  \times \left\langle l_3\left|e^{i\frac{K}{\kbar}\cos
{\hat \theta}}\right|l_4 \right\rangle \left\langle
l_3'\left|e^{i\frac{K}{\kbar}\cos {\hat
\theta}}\right|l_4'\right\rangle ^*\, .
 \label{Kdef}
\end{eqnarray}
Furthermore, we write down explicitly the matrix elements of
${\cal X}$, Eq.~(\ref{Kdef}), as
\begin{equation}
{\cal X}=\int\int\int\int \frac{d\theta _1}{2\pi} \frac{d\theta
_1'}{2\pi}\frac{d\theta _2}{2\pi}\frac{d\theta _2'}{2\pi} \exp
\left[\frac{i}{\kbar}\left(S_{kin}+S_{p}\right)\right] \, ,
\label{Xangle}
\end{equation}
\noindent where $S_{kin}$ and $S_p$ are defined as
\begin{eqnarray}
S_{kin}   &  =  &   -\theta_1
\left(l_1-l_2\right)+\theta_1'\left(l_1'-l_2'\right)
\nonumber\\
  &    &   - \theta_2\left(l_3-l_4\right)+
\theta_2'\left(l_3'-l_4'\right) \label{Skin}
\end{eqnarray}
and
\begin{equation}
S_{p} = K\left( \cos\theta_1 -\cos\theta_1'+\cos\theta_2
-\cos\theta_2' \right) \, , \label{Sp}
\end{equation}
respectively. For the discussions below, we make the following
change of variables:
\begin{eqnarray}
  &    &  \phi=\left(\theta_1-\theta_1'\right)-
\left(\theta_2-\theta_2'\right) \, , \nonumber \\
  &    &   \delta l_1=\frac{l_2+l_3'}{2}-\frac{l_3+l_2'}{2} \, ,
\quad
\delta\theta_1=\frac{\theta_1+\theta_1'}{2}+\frac{\theta_2+\theta_2'}{2}
\, ,
\nonumber\\
  &    &   \delta l_2=\frac{l_1+l_1'}{2}-\frac{l_4+l_4'}{2} \, ,
\nonumber\\
  &    &   \delta\theta_2=-\frac{1}{2}
\left[\left(\theta_1-\theta_1'\right)+
\left(\theta_2-\theta_2'\right)\right]\, . \label{notation2}
\end{eqnarray}
Then $S_{kin}$ may be rewritten as:
\begin{widetext}
\begin{eqnarray}
S_{kin}
  &  =  &  \frac{\phi}{2}\left[\left(\frac{l_2+l_3'}{2}+\frac{l_3+l_2'}{2}\right)-
\left(\frac{l_1+l_1'}{2}+\frac{l_4+l_4'}{2}\right)\right]+
\delta\theta_1 \, \delta l_1+ \delta\theta_2 \, \delta l_2
\nonumber\\
  &    &
-\frac{1}{2}\left(\frac{\theta_1+\theta_1'}{2}-\frac{\theta_2+\theta_2'}{2}\right)
\left\{\left[\left(l_1-l_1'\right)+\left(l_4-l_4'\right)\right]-
\left[\left(l_2-l_3'\right)+\left(l_3-l_2'\right)\right]
\right\}\nonumber\\
  &    &
-\frac{\delta\theta_1}{2}\left[\left(l_1-l_1'\right)-\left(l_4-l_4'\right)\right
] -
\frac{\delta\theta_2}{2}\left[\left(l_2-l_3'\right)-\left(l_3-l_2'\right)\right]
\, . \label{Skin1}
\end{eqnarray}
On the other hand, with the semiclassical approximation (i.e.
$\theta_{1,2}\approx \theta_{1,2}'$) taken into account, $S_{p}$
may be written as:
\begin{equation}
S_{p}\approx -2K\, \delta\theta_2 \,
\sin\left(\frac{\delta\theta_1}{2}\right)
\cos\left[\frac{1}{2}\left(\frac{\theta_1+\theta_1'}{2}-\frac{\theta_2+\theta_2'
}{2}\right)\right] -K\, \phi\,
\cos\left[\frac{1}{2}\left(\frac{\theta_1+\theta_1'}{2}-\frac{\theta_2+\theta_2'
}{2}\right)\right] \, . \label{Sp1}
\end{equation}
\end{widetext}
\noindent For sufficiently large $K$,
$[(\theta_1+\theta_1')-(\theta_2+\theta_2')]/4$ may be regarded as
a random phase. Therefore, the exponent becomes self-averaging
over this phase. Moreover, $\delta\theta_{1,2}\approx 0$. Finally,
we employ the conventional hydrodynamic approximation, i.e. take
into account the leading term in the $K\phi$-expansion only, to
arrive at
\begin{eqnarray}
  &    &   \exp \left(\frac{iS_{p}}{\kbar}\right) \rightarrow
\left\langle
 \exp \left(\frac{iS_{p}}{\kbar}\right)\right\rangle
\nonumber\\
  &  \approx  &   \left\langle\exp\left\{-\frac{iK\,
\delta\theta_1\, \delta\theta_2}{\kbar}\,
\cos\left[\frac{1}{2}\left(\frac{\theta_1+\theta_1'}{2}-\frac{\theta_2+\theta_2'
}{2}\right)\right]\right\}\right\rangle
\nonumber\\
  &  =  &   J_0
\left(\frac{K\,\delta\theta_1\delta\theta_2}{\kbar} \right) \, .
\label{ExpSp}
\end{eqnarray}

Let us insert Eqs.~(\ref{Xangle}), (\ref{Skin1}), and
(\ref{ExpSp}), as well as Wigner transform of ${\cal A}$, ${\cal
B}$ [cf. Eq.~(\ref{DWig})], and ${\cal C}_0$ [cf.
Eq.~(\ref{CWig})] into Eq.~(\ref{Iq}), and integrate out
$(l_1-l_1')/\kbar$, $(l_4-l_4')/\kbar$, $(l_2-l_3')/\kbar$,
$(l_3-l_2')/\kbar$, $(l_+-l_-)/\kbar$, $(l_+'-l_-')/\kbar$ and
$\phi/\kbar$. As a result, we find that the semiclassical
approximation for ${\cal I}_{q}$ is
\begin{eqnarray}
{\cal I}_{q} \left(l,\theta;l',\theta'\right)  &  =  &  {\hat
{\cal V}}\, \bigg[ {\cal A}_{\rm W}\left(l,\theta;l''+\frac{\delta
l_2}{2},\theta''+\frac{\delta\theta_1}{2}\right)  \label{1loopCC}\\
  &    &   \times {\cal B}_{\rm W}\left(l''-\frac{\delta
l_2}{2},-\theta''+\frac{\delta\theta_1}{2};l',\theta'\right)
\bigg] \, , \nonumber
\end{eqnarray}
where the vertex operator ${\hat {\cal V}}$ is an integral
operator: ${\hat {\cal V}} \, f (l,\theta;l',\theta';
l'',\theta'';\delta l_2, \delta \theta_1 ) \rightarrow ({\hat
{\cal V}} f)\, (l,\theta;l',\theta')$ (note that the variables:
$l'',\, \theta'',\, \delta l_2,\, $and $ \delta \theta_1$ in the
function $f$ are integrated out) and is defined as
\begin{eqnarray}
  &    &   \left({\hat {\cal V}} f \right)\, (l,\theta;l',\theta')
\label{interactionV}\\
  &  \equiv  &   \int\frac{dl''d\theta''}{2\pi\kbar}
\int\frac{d\delta l_1d\delta\theta_1}{2\pi\kbar} \int\frac{d\delta
l_2d\delta\theta_2}{2\pi\kbar}\, {\cal X}
(\delta l_1, \delta \theta_1 ;\delta l_2, \delta \theta_2)\nonumber\\
  &    &   \times {\cal C}_0\left(l''+\frac{\delta
l_1}{2},\theta''-\frac{\delta\theta_2}{2};l''-\frac{\delta
l_1}{2},-\theta''-\frac{\delta\theta_2}{2}\right) \nonumber\\
  &    &   \times f\, \left(l,\theta;l',\theta'; l'',\theta'';\delta l_2, \delta \theta_1 \right)\nonumber
\end{eqnarray}
with
\begin{eqnarray}
  &    &  {\cal X}\left(\delta l_1,\,\delta\theta_1;\,\delta
l_2,\,\delta\theta_2\right)\nonumber\\
  &  =  &   J_0
\left(\frac{K\,\delta\theta_1\delta\theta_2}{\kbar} \right)\,
\exp\left[ \frac{i}{\kbar}\left(\delta l_1\delta\theta_1+\delta
l_2\delta \theta_2\right) \right] \, .
 \label{Xres}
\end{eqnarray}
In Eq.~(\ref{1loopCC}) we introduced the following notations
\begin{eqnarray}
l''  &  \equiv  &   \frac{1}{2}\,\left(
\frac{l_2+l_3'}{2}+\frac{l_3+l_2'}{2}\right)  \, ; \nonumber\\
\theta''  &  \equiv  &   \frac{1}{2}\,\left(
\frac{\theta_1+\theta_1'}{2}-\frac{\theta_2+\theta_2'}{2}\right)
\, ,
 \label{notation3}
\end{eqnarray}
while the subscript ${\rm W}$ stands for Wigner transformation.

The minimal quantum wave packet posses the uncertainty: $\delta
l_1\delta\theta_1\, \sim\kbar $ and $\delta l_2\delta\theta_2 \,
\sim \kbar$, as indicated in Eq.~(\ref{Xres}). Moreover, $\delta
\theta_1\delta\theta_2 \, \sim \kbar/K$. Such quantum wave packet
determines the initial scale of the deviation of two nearby
trajectories involved in the Cooperon [cf. Eq.~(\ref{EhrTimeS})].

\subsubsection{Interaction vertex at the semiclassical level}
\label{vertexexact}

Applying the general expression Eq.~(\ref{1loopCC}) to
Eq.~(\ref{oneloopexact}), we find
\begin{eqnarray}
  &    &   \delta {\cal D}\left(l,\theta;l',\theta'\right)  \label{exactvertex4}\\
  &  =  &   {\hat {\cal V}}\, \bigg [{\cal
D}_0\left(l,\theta;l''+\frac{\delta
l_2}{2},\theta''+\frac{\delta\theta_1}{2}\right)\,\left(
\overleftarrow{P}_T-1\right)\nonumber\\
  &    &  \times \left( \overrightarrow{P}-1\right)\, {\cal
D}_0\left(l''-\frac{\delta
l_2}{2},-\theta''+\frac{\delta\theta_1}{2};l',\theta'\right)\bigg
]\, . \nonumber
\end{eqnarray}
Remarkably, this exact vertex $\delta {\cal D}$ is obtained
without introducing any explicit regularization. In the next
subsection we will show that it leads to the weak--localization
correction to diffusion constant, which is similar to earlier
findings for ballistic quantum dots.\cite{AL96} In
Appendix~\ref{ALRegularization} we  show that  it is possible to
introduce some artificial quantum disorder to QKR and the results
obtained in the present work may be reproduced following the
formalism of Ref.~\onlinecite{AL96}.

\subsection{Weak localization correction to diffusion constant}
\label{WLC}

In this part, we show that  the exact one--loop quantum
correction, Eq.~(\ref{exactvertex4}), may be cast into the
conventional Hikami box structure with an additional, factor due
to the Lyapunov region. As a result, the one-loop correction
affects the diffusion equation through a frequency--dependent
renormalization of the  diffusion coefficient: $D(\omega) = D_{cl}
+\delta D(\omega)$. That is,
\begin{equation}
\left[-i\omega -D(\omega)\nabla_l ^2\right]{\cal D}_0 = \kbar\,
\delta\left(l-l'\right). \label{ModifClDiff}
\end{equation}
We calculate then the  quantum correction $\delta D(\omega)$ and
find how it affects the angular momentum  dispersion.

\subsubsection{Effects of the Lyapunov instability on the interaction vertex}
\label{diffusivevertex}

Since our aim is to describe the long--time phenomena, we expect
that the typical scale of the angular momentum dispersion, say
$L_H$, is large: $L_H\gg K$. Indeed the angular momentum is
randomly spread in the interval $\in [-K,K]$ in a single kick. It
is thus natural to expect much broader  distribution after many
kicks. This consideration justifies  the expansion with respect to
$K\, \nabla_l$ (hydrodynamic approximation). With such
approximation, the exact vertex can be cast into the Hikami box.

With the help of the identity: $ \overrightarrow{P}\, {\cal
D}_0={\cal D}_0\, \overleftarrow{P}_T $\,, which is proven in
Appendix~\ref{relation}, Eq.~(\ref{exactvertex4}) may be rewritten
as
\begin{eqnarray}
  &    &   \delta {\cal D}\left(l,\theta;l',\theta'\right)  \label{exactvertex4eqv}\\
  &  =  &   {\hat {\cal V}}\, \bigg \{ \left[\left(
\overrightarrow{P}-1\right)\, {\cal
D}_0\left(l,\theta;l''+\frac{\delta
l_2}{2},\theta''+\frac{\delta\theta_1}{2}\right) \right]\nonumber\\
  &    &  \times \left[{\cal D}_0\left(l''-\frac{\delta
l_2}{2},-\theta''+\frac{\delta\theta_1}{2};l',\theta'\right)\,\left(
 \overleftarrow{P}_T-1\right) \right]\bigg \}\, . \nonumber
\end{eqnarray}

Since ${\cal X}\, {\cal C}_0$ has no dependence on $(l'',\,
\theta'')$,  Eq.~(\ref{exactvertex4eqv}) contains
\begin{eqnarray}
I  &  =  &  \int\frac{dl''d\theta''}{2\pi\kbar}\, {\cal
D}_0\left(l,\theta;l''+\frac{\delta
l_2}{2},\theta''+\frac{\delta\theta_1}{2}\right)\nonumber\\
  &    &   \times {\cal D}_0\left(l''-\frac{\delta
l_2}{2},-\theta''+\frac{\delta\theta_1}{2};l',\theta'\right) \, .
\label{twodiffusonintegral}
\end{eqnarray}

To proceed further, we employ the following relation:
\begin{equation}
{\cal D}_0\left(l,\theta;l',\theta'\right)= {\cal
D}_0\left(l',-\theta';l,-\theta\right) \, ,
\label{DDtimereversalrela}
\end{equation}
which reflects the time-reversibility and its derivation is given
in Appendix~\ref{relation}. The remaining procedure is fully
analogous to calculation of the  Cooperon developed in
Sec.~\ref{coop}. The only difference is in the boundary
conditions. In fact, the angular deviation of two travelling
nearby trajectories, $\delta\theta$  reaches a classical size
$\left|\delta\theta''\right|\lesssim 1$ at some point in the phase
space, say $(l_{n_c}\,,\theta_{n_c})$\,. The later evolutions are
independent. That is, the two Diffusons become self-averaging over
the (random) paths connecting two remote ends, resulting in the
factorization of the  two averaged Diffusons as
\begin{eqnarray}
  &    &  \bigg [ {\cal D}_0\left(l''+\frac{\delta
l_2}{2},-\theta''-\frac{\delta\theta_1}{2};l,-\theta\right)
\nonumber\\
  &    &   \times {\cal D}_0\left(l''-\frac{\delta
l_2}{2},-\theta''+\frac{\delta\theta_1}{2};l',\theta'\right)
\bigg ]\nonumber\\
  &  \rightarrow  &    \left \langle {\cal
D}_0\left(l_1+\frac{\delta
l''}{2},-\theta_1-\frac{\delta\theta''}{2};l,-\theta\right)\right
\rangle \nonumber\\
  &    &   \times \left\langle{\cal D}_0\left(l_1-\frac{\delta
l''}{2},-\theta_1+\frac{\delta\theta''}{2};l',\theta'\right)\right
\rangle\, . \label{DDBC}
\end{eqnarray}
Taking this boundary condition into account, we obtain:
\begin{eqnarray}
I  &  =  &  {\cal W}_{\rm D}(2\omega)\, \int
\frac{dl_1d\theta_1}{2\pi\kbar}\, \left\langle{\cal
D}_0\left(l,\theta;l_1+\frac{\delta
l''}{2},\theta_1+\frac{\delta\theta''}{2}\right) \right\rangle\nonumber\\
  &    &  \times \left\langle{\cal D}_0\left(l_1-\frac{\delta
l''}{2},-\theta_1+\frac{\delta\theta''}{2};l',\theta'
\right)\right\rangle \, . \label{DDWLC}
\end{eqnarray}
Note that the two intermediate angular momenta deviate as $\delta
l''/2$.  Such  deviation is unimportant because the distribution
with respect to the angular momentum fluctuates over the large
scale $L_H\gg K\gg \delta l''/2$\,. In Eq.~(\ref{DDWLC}) ${\cal
W}_{\rm D}(\omega)$ is the same as ${\cal W}_{\rm C}(\omega)$
except that $t^{\rm C}$ is replaced by (with the logarithmic
accuracy)\,:
\begin{equation}
t^{\rm
D}=\frac{1}{\lambda}\left|\ln\frac{1}{\sqrt{\delta\theta^2+\delta
l^2/K^2}}\right| \, , \label{EhrTimeUs}
\end{equation}
where $\delta\theta$ is the initial angular separation of two
nearby trajectories involved in the Diffuson side of the Lyapunov
region. It is determined by the minimal quantum wave packet, i.e.,
by ${\cal X}$ [see Eq.~(\ref{Xres})]\,. We then substitute
Eqs.~(\ref{interactionV}) and (\ref{DDWLC}) into
Eq.~(\ref{exactvertex4eqv}), and restore the operator under the
average. As a result, we obtain:
\begin{widetext}
\begin{eqnarray}
  &    &   \delta {\cal D}\left(l,\theta;l',\theta'\right)
  \label{diffusivetvertex4}\\
  &   =   &   {\cal
V}\, \int \frac{dl_1d\theta_1}{2\pi\kbar}
  \bigg \{ \left\langle{\cal
D}_0\left(l,\theta;l_1+\frac{\delta
l''}{2},\theta_1+\frac{\delta\theta''}{2}\right)\, \left(
\overleftarrow{P}_T-1\right) \right\rangle \,
\left\langle \left(
\overrightarrow{P}-1\right)\,{\cal D}_0\left(l_1-\frac{\delta
l''}{2},-\theta_1+\frac{\delta\theta''}{2};l',\theta'
\right)\right\rangle \bigg \}\, , \nonumber
\end{eqnarray}
\end{widetext}
where
\begin{eqnarray}
{\cal V}  &  \equiv  &   \int\frac{d\delta
l_1d\delta\theta_1}{2\pi\kbar} \int\frac{d\delta
l_2d\delta\theta_2}{2\pi\kbar}\,{\cal W}_{\rm D}(2\omega)\, {\cal
X} (\delta l_1, \delta \theta_1 ;\delta l_2, \delta \theta_2)
\nonumber\\
   &  \times    &   {\cal C}_0\left(l''+\frac{\delta
l_1}{2},\theta''-\frac{\delta\theta_2}{2};l''-\frac{\delta
l_1}{2},-\theta''-\frac{\delta\theta_2}{2}\right)  \, .
\label{Vrepresentation}
\end{eqnarray}

Here ${\cal V}$ may be regarded as the renormalized interaction
strength. To further calculate it, we substitute
Eq.~(\ref{CoopRenor}) into Eq.~(\ref{Vrepresentation}). As a
result, ${\cal V}$ is found to be
\begin{equation}
{\cal V} =\Gamma(\omega)\!
 \int \, \frac{d \varphi}{2\pi} \frac{\kbar}{-i\omega+D_{cl}\varphi^2}\,\, ,
                                                       \label{VRenormalization}
\end{equation}
where
\begin{eqnarray}
\Gamma(\omega)   &  \equiv  &   \int\frac{d\delta
l_1d\delta\theta_1}{2\pi\kbar} \,\int\frac{d\delta
l_2d\delta\theta_2}{2\pi\kbar} \, {\cal W}_{\rm
C}\left(2\omega\right)\, {\cal W}_{\rm D}\left(2\omega\right)\,
{\cal X}
\nonumber\\
  &  =  &   \int\frac{d\delta l_1d\delta\theta_1}{2\pi\kbar}
\,\int\frac{d\delta l_2d\delta\theta_2}{2\pi\kbar}
\nonumber\\
  &    &  \times J_0
\left(\frac{K\,\delta\theta_1\delta\theta_2}{\kbar} \right)\,
\exp\left[ \frac{i}{\kbar}\left(\delta l_1\delta\theta_1+\delta
l_2\delta
\theta_2\right) \right] \nonumber\\
  &    &  \times \exp \left[\left(2i\omega -\frac{2\omega
^2\lambda _2}{\lambda ^2}\right) \left|z_1+z_2\right| \right]
\label{Gam}
\end{eqnarray}
with
\begin{equation}
z_1=\ln \sqrt{\delta \theta_1^2+\delta l_2^2/K^2} \, ,\quad
z_2=\ln \sqrt{\delta \theta_2^2+\delta l_1^2/K^2} \, .
\label{z1z2}
\end{equation}
Rescaling $\delta \theta _{1,2}$ and $\delta l_{1,2}$ as
\begin{equation}
\delta \theta _{1,2}  \rightarrow  \delta \theta _{1,2} /
\sqrt{\kbar/K} \, ,\quad \delta l _{1,2}  \rightarrow  \delta l
_{1,2} / \sqrt{\kbar K} \label{rescale}
\end{equation}
leads to
\begin{equation}
\Gamma (\omega) =\exp \left(4i\omega t_E -\frac{4\omega ^2 \lambda
_2 t_E}{\lambda ^2}\right) \, F(\omega ) \, ,\label{GammaRescale}
\end{equation}
where $F(\omega)$ is
\begin{eqnarray}
F(\omega) &   =  &   \int\frac{d l_1d \theta_1}{2\pi }
\,\int\frac{d l_2d \theta_2}{2\pi } \,
 J_0 \left( \theta_1
\theta_2  \right)  \nonumber\\
    &       &
\times \exp\left[
 i\left(l_1 \theta_1+  l_2
\theta_2\right) \right] \nonumber\\
  &    &  \times \exp \left[\left(2i\omega -\frac{2\omega
^2\lambda _2}{\lambda ^2}\right) \left|\widetilde{z}_1 +
\widetilde{z}_2\right| \right] \, ,  \label{F} \\
\widetilde{z}_1
   &       =      &       \ln \sqrt{ \theta_1^2+ l_2^2 } \, ,
   \quad \widetilde{z}_2      =       \ln
\sqrt{ \theta_2^2+ l_1^2} \, . \nonumber
\end{eqnarray}
Here $t_E=\left(t^{\rm C}+t^{\rm D}\right)/2$, which is
Eq.~(\ref{ehrenfest}).  Since $\widetilde{z}_{1,2} \sim 1$, in the
limit $\omega,~\omega\lambda_2/\lambda^2\ll 1$, the last exponent
in Eq.~(\ref{F}) may be considered to be $1$. As a result,
$F(\omega) =1$. Thus, we obtain:
\begin{equation}
\Gamma (\omega) =\exp \left(4i\omega t_E -\frac{4\omega ^2 \lambda
_2 t_E}{\lambda ^2}\right). \label{Gamma1}
\end{equation}

We point out that the position of minimal wave packet--Hikami
box--can {\it not} be exactly located within the Lyapunov region.
Indeed, this is reflected in the fact that the total duration
within the Lyapunov region, i.e., $4 t_E$ actually does {\it not}
depend on the exact boundary between Cooperon and Diffuson. Such
feature originates from the chaotic nature of the classical motion
in the phase space. At each full travel, the initial deviation,
$\delta l_{1,2}$ ($\delta \theta_{1,2}$) with respect to the
reference trajectory expands in the backward (forward) time
direction. Eventually $\delta l_1\, \delta \theta_1$ $(\delta
l_2\, \delta \theta_2)$ reaches some classical action $K$ (the
typical scale of the classical action)\,. Therefore, the total
duration is
\begin{equation}
 4t_E = \frac{1}{\lambda}\ln \frac{K}{\delta l_1\delta \theta_1}+
\frac{1}{\lambda}\ln \frac{K}{\delta l_2\delta \theta_2} \, .
\label{tEdecomposition}
\end{equation}
Taking into account the uncertainty relation: $\delta l_1\delta
\theta_1\approx\delta l_2\delta \theta_2\approx \kbar\,$, we find
the total duration to be
$ 4t_E=  4\lambda^{-1}\ln \sqrt{K/\kbar} $\,.

\subsubsection{Frequency-dependent diffusion coefficient}
\label{WLCdifcoe}

The renormalized interaction vertex, Eq.~(\ref{diffusivetvertex4})
may be further cast into the conventional (diffusive) Hikami box
upon further simplifications. As discussed above,  the
hydrodynamic expansion may be performed because of $K/L_H \ll 1$.
This allows the further simplification of
Eq.~(\ref{diffusivetvertex4}). One finds to the first order in the
hydrodynamic expansion
\begin{widetext}
\begin{eqnarray}
  &    &   \left\langle\overrightarrow{P} \,{\cal
D}_0\left(l_1-\frac{\delta
l''}{2},-\theta_1+\frac{\delta\theta''}{2};l',\theta'
\right) \right\rangle \nonumber\\
  &  =  &  \left\langle {\cal D}_0\left(l_1-\frac{\delta
l''}{2}+K\sin
\left[\theta_1-\frac{\delta\theta''}{2}+l_1-\frac{\delta
l''}{2}\right],-\left[\theta_1-\frac{\delta\theta''}{2}+l_1-\frac{\delta
l''}{2}\right];l',\theta'
\right) \right\rangle\nonumber\\
  &  \approx  &   \left\langle \left[1+K\sin
\left(\theta_1-\frac{\delta\theta''}{2}+l_1-\frac{\delta
l''}{2}\right) \nabla_{l_1} \right]\,
{\cal D}_0\left(l_1-\frac{\delta l''}{2},
-\left[\theta_1-\frac{\delta\theta''}{2}+l_1-\frac{\delta
l''}{2}\right];l',\theta' \right) \right\rangle\nonumber\\
  &  \approx  &   \left[1+K\sin
\left(\theta_1-\frac{\delta\theta''}{2} +l_1-\frac{\delta l''}{2}
\right) \nabla_{l_1} \right]\, \left\langle {\cal
D}_0\left(l_1-\frac{\delta l''}{2},
-\left[\theta_1-\frac{\delta\theta''}{2}+l_1-\frac{\delta
l''}{2}\right];l',\theta' \right) \right\rangle\, .
\label{Pexpansion}
\end{eqnarray}
The last line results from the fact that $\langle {\cal
D}_0\rangle$ has weaker dependence on the angle compared to the
sinusoidal term for sufficiently large $K$. Similarly,
\begin{eqnarray}
  &    &   \left\langle{\cal D}_0\left(l,\theta;l_1+\frac{\delta
l''}{2},\theta_1+\frac{\delta\theta''}{2}
\right) \, \overleftarrow{P}_T \right\rangle\, \nonumber\\
  &  \approx  &   \left[1+K\sin
\left(\theta_1+\frac{\delta\theta''}{2}+l_1+\frac{\delta l''
}{2}\right) \nabla_{l_1} \right]\, \left\langle {\cal
D}_0\left(l,\theta;l_1+\frac{\delta l''}{2},\,
\theta_1+\frac{\delta\theta''}{2}+l_1-\frac{\delta l''}{2}
\right)\right\rangle
\, . \label{PTexpansion}
\end{eqnarray}
On the other hand, by shifting the overall angle factor,
\begin{eqnarray}
  &    &   \int \frac{dl_1d\theta_1}{2\pi\kbar}\, \left\langle
{\cal D}_0\left(l_1-\frac{\delta l''}{2},
-\left[\theta_1-\frac{\delta\theta''}{2}+l_1-\frac{\delta
l''}{2}\right];l',\theta' \right) \right\rangle\, \left\langle
{\cal D}_0\left(l,\theta;l_1+\frac{\delta l''}{2},\,
\theta_1+\frac{\delta\theta''}{2}+l_1-\frac{\delta l''}{2}
\right)\right\rangle \nonumber\\
  &  =  &   \int \frac{dl_1d\theta_1}{2\pi\kbar}\,
\left\langle{\cal D}_0\left(l_1+\frac{\delta
l''}{2},-\theta_1+\frac{\delta\theta''}{2};l',\theta' \right)
\right\rangle\, \left\langle {\cal
D}_0\left(l,\theta;l_1+\frac{\delta l''}{2},\,
\theta_1+\frac{\delta\theta''}{2} \right)\right\rangle \, .
\label{DDangleshift}
\end{eqnarray}
This arises from the uniform distribution with respect to the
common angle in the product of the two averaged Diffusons.

We then substitute these two expansions: Eqs.~(\ref{Pexpansion})
and (\ref{PTexpansion}), as well as Eq.~(\ref{DDangleshift}) into
Eq.~(\ref{diffusivetvertex4}). For sufficiently large $K$, the
sinusoidal term is quasi-random and may be averaged over the
angular region $[0\,,2\pi]$. As a result, the linear term in the
hydrodynamic expansion does not survive upon this averaging and
the second order term must be kept.  Finally we obtain:
\begin{eqnarray}
\delta {\cal D}\left(l,\theta;l',\theta'\right)  &  =  &  {\cal
V}\, \int \frac{dl_1d\theta_1}{2\pi\kbar}\, K^2 \left\langle\sin
\left(\theta_1-\frac{\delta\theta''}{2}\right)\, \sin
\left(\theta_1+\frac{\delta\theta''}{2}\right)\right\rangle \,
\nabla_{l_1'}\, \nabla_{l_1''}\nonumber\\
  &  \times  &   \left[\left\langle{\cal
D}_0\left(l,\theta;l_1'+\frac{\delta
l''}{2},\theta_1+\frac{\delta\theta''}{2}\right)\right\rangle \,
\left\langle {\cal D}_0 \left(l_1''-\frac{\delta
l''}{2},-\theta_1+\frac{\delta\theta''}{2};l',\theta'
\right)\right\rangle\right] \bigg |_{l_1'=l_1''=l_1}\,
.\label{divertex}
\end{eqnarray}
\end{widetext}
\noindent Since $\delta\theta'' \ll 1 \sim \theta_1$,
\begin{eqnarray}
  &    &   K^2 \left\langle\sin
\left(\theta_1-\frac{\delta\theta''}{2}\right)\, \sin
\left(\theta_1+\frac{\delta\theta''}{2}\right)\right\rangle
\nonumber\\
  &  \approx  &   K^2 \left\langle \sin^2 \theta_1\right\rangle =
2D_{cl} \, . \label{sinsin}
\end{eqnarray}
On the other hand, the Diffuson is smooth over the scale $\sim K$,
hence we may also drop out $\delta l''/2$ in Eq.~(\ref{divertex}).
Finally $\delta {\cal D}$ is cast into
\begin{eqnarray}
\delta {\cal D}\left(l,\theta;l',\theta'\right) = 2{\cal V}\, \int
\frac{dl_1d\theta_1}{2\pi\kbar}\, D_{cl}\, \nabla_{l_1'}\,
\nabla_{l_1''}
\label{WDLDiffusonSim} \\
\times \left[\left\langle{\cal D}_0\left(l,\theta;l_1'
, \theta_1
\right) \right\rangle \, \left\langle {\cal D}_0\left(l_1''
,-\theta_1
;l',\theta' \right)\right\rangle\right]\, |_{l_1'=l_1''=l_1}\,
.\nonumber
\end{eqnarray}
The diffusion coefficient appearing in Eq.~(\ref{WDLDiffusonSim})
is $D_{cl}=K^2/4$. We argue that if more kicks are reserved for
the Hikami box, i.e., $({\hat P}-1)^n\, , n  >   1$, then the
diffusion coefficient will acquire the same higher order
corrections as Eq.~(\ref{Dcl}).

With the angle averaged out diffusion is retrieved. As a result,
the one-loop correction, Eq.~(\ref{WDLDiffusonSim}), is simplified
as
\begin{eqnarray}
\delta {\cal D}_0\left(l,l'\right)=\kbar^{-1} \int dl_1\,{\cal V}
D_{cl}\,
\left[\nabla_{l_1'}^2+\nabla_{l_1''}^2\right]\nonumber\\
\times \left[ {\cal D}_0\left(l,l_1'\right) {\cal
D}_0\left(l_1'',l'\right) \right] |_{l_1'=l_1''=l_1} \, .
\label{WDLDiffuson}
\end{eqnarray}
Denote the Fourier transform of ${\cal D}_0\left(l,l'\right)$,
with respect to $l-l'$ as ${\cal D}_0(\varphi;\omega)$.
Substituting Eq.~(\ref{VRenormalization}) into
Eq.~(\ref{WDLDiffuson}) we find the Fourier transform one-loop
correction:
\begin{equation}
\delta {\cal D}_0(\varphi;\omega) =
\frac{\kbar\Gamma(\omega)D_{cl}\, \varphi^2}{\left(-i\omega
+D_{cl}\, \varphi^2\right)^2} \int \frac{d\phi}{\pi}
\frac{1}{-i\omega + D_{cl}\, \phi ^2} \, ,
\label{DDclKW}
\end{equation}
which leads to the one-loop quantum correction to the diffusion
coefficient as
\begin{equation}
\delta D(\omega) =-{\kbar D_{cl}\over \pi}\, \Gamma(\omega)\!
 \int \!\! \frac{d \phi}{-i\omega+D_{cl}\phi^2}\,\, .
                                                       \label{DDclKW}
\end{equation}

\subsubsection{Dispersion function}
\label{momdis}

One may express the time evolution of the angular momentum
dispersion as:
$ \delta \langle l^2(t)\rangle \equiv \langle (l(t)-l(0) )^2
\rangle $
in terms of the frequency-dependent diffusion coefficient. In
fact, by averaging over the angle, we may write $\delta \langle
l^2(t)\rangle$ as
\begin{eqnarray}
\delta\left\langle l^2(t)\right\rangle
  &  =  &   \sum_l \, \left(l-l'\right)^2 \left[ {\cal D}_0
\left(l, l';
t\right)- {\cal D}_0 \left(l, l'; 0\right)\right] \nonumber\\
  &  =  &   -\frac{\partial^2}{\partial \varphi^2}\, \left[{\cal
D}_0 \left(\varphi; t\right)- {\cal D}_0 \left(\varphi;
0\right)\right] |_{\varphi\rightarrow 0} \, .
\label{TimeCorrClDiff}
\end{eqnarray}
Substituting Eq.~(\ref{ModifClDiff}) into it, we obtain:
\begin{equation}
\delta\left\langle l^2(t) \right\rangle =\!
\int\limits_{-\infty}^{\infty}\!\! {d\omega\over \pi}\,
\frac{1-e^{-i\omega t}}{\omega^2}\,\,  D(\omega)\, .
                                                 \label{TimeCorrDC}
\end{equation}

For sufficiently large $K$ one may ignore the fluctuation of
$\lambda$, i.e. put $\lambda_2=0$. Consequently, in the leading
order in $\kbar$ the momentum dispersion is found to be:
\begin{equation}
\delta\langle l^2(t)\rangle =2D_{cl}t - \frac{8\kbar
\sqrt{D_{cl}}}{3\sqrt \pi}\,\theta (t- 4t_E) \left(t-
4t_E\right)^{3/2} ,
                                           \label{result}
\end{equation}
where $\theta(t)$ is the step function (long-dashed line on
Fig.~\ref{fig1}). The singularity at $t=4t_E$ is rounded by the
Ehrenfest time fluctuations--arising from finite $\lambda_2$ (full
line on Fig.~\ref{fig1}). Substituting Eqs.~(\ref{Gamma1}) and
(\ref{DDclKW}) into Eq.~(\ref{TimeCorrDC}), we arrive at
\begin{eqnarray}
   &   &   \delta\langle l^2(t)\rangle -2D_{cl}t = -\frac{4\kbar \sqrt
{D_{cl}}}{3\pi} \left(\delta t_E\right)^{3/2}
\label{convolution}\\
   &   \times    &   \int_{-\infty}^\infty \, d\tau \theta\left(\frac{t-4
t_E}{\delta t_E}-\tau\right) \left(\frac{t-4 t_E}{\delta
t_E}-\tau\right)^{3/2} \, e^{-\frac{\tau^2}{16}} \, , \nonumber
\end{eqnarray}
where $\delta t_E = \sqrt {\lambda_2 t_E/ \lambda^2}$. As a
result,
\begin{equation}
\delta\langle l^2(t)\rangle =2D_{cl}t -\frac{\Gamma
\left(\frac{5}{4}\right)\kbar}{3\pi/64 }  \sqrt{D_{cl}}\, (\delta
t_E)^{3/2} f\left({4t_E-t\over \delta t_E}\right)\, ,
                                            \label{smalltime}
\end{equation}
where $f(0)=1$ and
\begin{eqnarray}
f(x)=\bigg \{
\begin{array}{ll}
\frac{8\sqrt {2}\Gamma\left(\frac{7}{2}\right)}{\Gamma
\left(\frac{5}{4}\right)} \,
x^{-5/2}\, e^{-x^2/16}\, & \quad {\rm for}\,\,\,\, \, x\gg 1 \, . \\
\frac{1}{8 \Gamma \, \left(\frac{5}{4}\right)} \, (-x)^{3/2} \, &
\quad {\rm for}\,\,\,
-x\gg 1 \, .\\
\end{array}
 \label{fasymptotics}
\end{eqnarray}
This result completes the calculations of the one--loop
weak--localization correction.

\section{Weak Dynamical Localization of QKR with Broken Time-Reversal Symmetry}
\label{QuantumCorr2Loop}

To exploit farther similarities and differences of the dynamical
localization of the QKR and the Anderson localization we discuss
here effects of breaking the time--reversal symmetry (TRS). In
case of Anderson localization in a random potential the TRS may be
broken, for example, by a static magnetic field. The latter
provides different phases for clock-wise and anti-clock-wise
propagating trajectories, destroying thus the systematic
interference correction discussed in Section II. It does not ruin,
though, the Anderson localization completely. Indeed, higher order
corrections (the minimum possible is the two--loop one) may be
interpreted as interference of trajectories travelling  the loops
in the same direction only (Diffuson only diagrams with no
Cooperons). The static magnetic field does not affect such
diagrams. As a result, the Anderson localization exists even in
this case, albeit with somewhat larger localization length.

It has been thus  of long interest to show that an analogous
phenomena exists for the QKR as well
\cite{Izrailev90,Izrailev86,TBS93,TTB04,MRBF04}. Our additional
motivation comes from consideration of the Lyapunov regime and its
sensitivity to  the TRS breaking. In particular, the one--loop
(TRS invariant) correction was found to be delayed by $4t_E$. Does
this time interval remains to be protected against perturbative
corrections in higher loop processes ? Is the delay time still the
same ? These questions are of particular interest if and when the
leading one--loop correction is destroyed by TRS breaking.

\begin{figure}[h]
  \centerline{\epsfxsize=3in\epsfbox{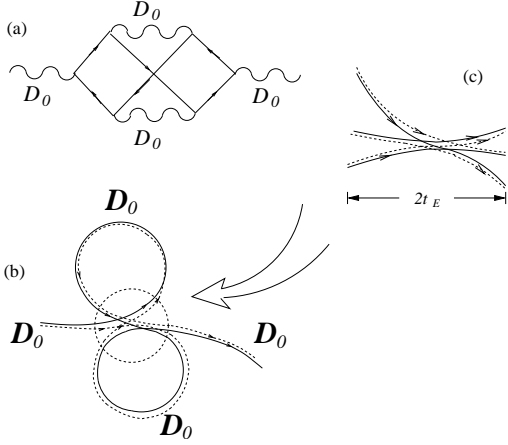}}
\caption{ The leading quantum correction to the density-density
correlator in the absence of the time-reversal symmetry: (a)
two-loop weak localization diagram; (b) its image in the momentum
space; (c) Hikami box.} \label{fig2loop}
\end{figure}

To answer these questions we investigate the model, described by
the following Hamiltonian
\begin{eqnarray}
{\hat H}           &               =           & \frac{{\hat
l}^2}{2} + K \sum_n \bigg [ \cos {\hat \theta}\,\,
\delta \left(t-2n\right) \nonumber\\
           &                          &                + \cos \left({\hat \theta}+\Phi\right)\, \delta
\left(t-(2n+1)\right) \bigg] \, . \label{SNHamiltonianrescale}
\end{eqnarray}
Here the time-reversal symmetry is broken for generic $\Phi$
except for $\Phi=0,\pi$\,. First, we analyze suppression  of the
Cooperon (and thus the one--loop diagram) arising from $\Phi$. We
then  calculate the two-loop correction. In contrast to the
one-loop correction, the two-loop one is robust against  $\Phi$
because it contains (among others) a diagram without the
Cooperons. If the TRS is broken, the long--time correction is
given by the  two-loop diagram depicted in Fig.~\ref{fig2loop}
\cite{SLA98,WLS99,TL03}. It differs from Fig.~\ref{fig2} in that
the two interfering paths propagate together in the {\em same}
direction, except inside the Hikami box, where they switch from
one to the other. Due to such geometry, the two-loop correction is
not sensitive to $\Phi$\,. It requires three successive travelling
through the Lyapunov region (each taken $2t_E$ time). We show thus
that the weak localization correction, given by this diagram, is
delayed by $6t_E$\, \cite{TL03}.

\subsection{Suppression of one-loop correction}
\label{smallphasebias}

In the modified KR model, Eq.~(\ref{SNHamiltonianrescale}), the
period is doubled and includes two kicking. The effective kicking
operator appearing in the one-step evolution operator ${\hat U}$
[cf. Eq.~(\ref{selfenergyPerron})] is replaced by
\begin{eqnarray}
\hat {\cal P}_{J}'           &               \equiv           &
\exp \left[ i\frac{K}{\kbar} \,\cos\hat\theta\right] \exp \left
[\frac{i\hat l
^2}{2\kbar}\right] \nonumber\\
           &                          &                \times \exp \left[ i\frac{K}{\kbar}
\,\cos\left(\hat\theta +\Phi\right)\right] \, . \label{U2}
\end{eqnarray}
In the angular momentum representation, the matrix elements read
as
\begin{eqnarray}
\left\langle l_+\left|{\hat {\cal P}}_{J}'\right|l_+'\right\rangle
           &               =           &
\sum_{l_1}\int\frac{d\theta_+}{2\pi}\int\frac{d\theta_1}{2\pi}
e^{-\frac{i}{\kbar}\theta_+\left(l_+-l_1\right)}e^{\frac{iK}{\kbar}\cos\theta_+}
\nonumber\\
           &                          &               \times e^{\frac{il_1^2}{2\kbar}}
e^{-\frac{i}{\kbar}\theta_1\left(l_1-l_+'\right)}e^{\frac{iK}{\kbar}\cos\left(
\theta_1+\Phi\right)}\, , \label{U2ele}
\end{eqnarray}
while its complex conjugation is
\begin{eqnarray}
\left\langle l_-\left|{\hat {\cal
P}}_{J}'\right|l_-'\right\rangle^*            & =           &
\sum_{l_1'}\int\frac{d\theta_-}{2\pi}\int\frac{d\theta_1'}{2\pi}
e^{\frac{i}{\kbar}\theta_-\left(l_--l_1'\right)}e^{-\frac{iK}{\kbar}\cos\theta_-
}
\nonumber\\
           &                          &               \times e^{-\frac{i l_1'^2}{2\kbar}}
e^{\frac{i}{\kbar}\theta_1'\left(l_1'-l_-'\right)}e^{-\frac{iK}{\kbar}\cos\left(
\theta_1'+\Phi\right)} \, . \label{U2ComCele}
\end{eqnarray}

First we investigate the effects of $\Phi$ on the   diffusive
parts of Diffusons  and  Cooperons. To simplify the discussion we
assume that $K$ is sufficiently large, implying  that the memory
about the angle is lost after a single kick. To find the
self--energy of the  Diffuson, we replace the kicking operator $J$
in Eq.~(\ref{selfenergydiff}) with ${\hat {\cal P}}_J'$\,. Then we
insert Eqs.~(\ref{U2ele}) and (\ref{U2ComCele}) into it, putting
$l_1=l_1'$\,. Consequently, the diffusive pole is retrieved as
$\theta_+-\theta_-=\theta_1-\theta_1'\equiv \kbar\varphi$\,, and
the self-energy of the Diffuson is
\begin{eqnarray}
     &          &     \langle l|{\hat {\cal P}}_J'|l'' \rangle \langle
l|{\hat {\cal P}}_J'|l'' \rangle ^* \nonumber\\
       &    =     &         \int \frac{d\varphi}{2\pi}\, J_0^2
\left(\frac{2K}{\kbar}\sin \frac{\kbar\varphi}{2}\right) \,
e^{i\varphi\left(l-l''\right)} \, . \label{selfenergydifftwokick}
\end{eqnarray}
This expression demonstrates the period doubling (Bessel function
is squared) and implies that the  Diffuson is not affected by
$\Phi$\,.

For the diffusive Cooperon, the self-energy is $\langle l|{\hat
{\cal P}}_J'|l'' \rangle \langle l''|{\hat {\cal P}}_J'|l \rangle
^*$\,. Inserting Eqs.~(\ref{U2ele}) and (\ref{U2ComCele}) into it,
and putting $l_1=l_1'$\,, we find that the Cooperon has diffusive
pole at
\begin{equation}
\theta_++\theta_-=\theta_1+\theta_1'\equiv \kbar\varphi \, ,
\label{Q}
\end{equation}
and the self-energy is
\begin{eqnarray}
\langle l|{\hat {\cal P}}_J'|l'' \rangle \langle l''|{\hat {\cal
P}}_J'|l \rangle^*  = \int \frac{d\varphi}{2\pi}\,
e^{i\varphi\left(l-l''\right)} \label{SigQ1}\\
\times  J_0\left(\frac{2K}{\kbar}\sin
\frac{\kbar\varphi}{2}\right)\,
J_0\left[\frac{2K}{\kbar}\sin\left(
\frac{\kbar\varphi}{2}+\Phi\right)\right]
 \nonumber \, . \nonumber
\end{eqnarray}
In the derivation above, we used the fact that for sufficiently
large $K$\,, $(\theta_+-\theta_-)/2$ and $(\theta_1-\theta_1')/2$
is quasi-random and the self-averaging may be performed. With the
Fourier transform with respect to time, the self-energy  leads to
the diffusive Cooperon of the form:
\begin{eqnarray}
           &                          &                \left\langle{\cal
C}_0(l,\theta;l',-\theta';\omega)\right\rangle \label{coopP}\\
           &               =           &                \kbar\int \frac{d\varphi}{2\pi}\frac{1}{-i\omega
+\frac{1}{2}\, D_{cl}
\left[\varphi^2+\left(\varphi+2\Phi/\kbar\right)^2\right]}
\nonumber
\end{eqnarray}
in the limit: $K\varphi,\, K\Phi/\kbar \ll 1$\,. Here
$|l-l'|\lesssim K,\, |\theta-\theta'|\sim 1$\,. According to
Eq.~(\ref{coopP}) the relaxation time of the Cooperon is
$\tau_\Phi=\left(D_{cl}\Phi^2/\kbar^2\right)^{-1}$\,. At
$\Phi\gtrsim \kbar/\sqrt{D_{cl}} \sim \kbar/K$\,, $\tau_\Phi\sim
1$\, the Cooperon is completely suppressed.

The propagation in the Lyapunov region involves deterministic
motion, which,  is also affected by  $\Phi$\,. Therefore the
Lyapunov exponent $\lambda$, as well as its fluctuations
$\lambda_2$ acquire some $\Phi$-dependence. However, since the
propagator has no dependence on the center of mass,  the
functional form of ${\cal W}_{\rm C,D}(2\omega)$ remains
unchanged. Thus, the procedure of Sec.~\ref{WLC} can be employed
to show that the dispersion function  is given by:
\begin{equation}
\delta \left\langle l^2(t)\right\rangle = 2D_{cl} t-
\frac{4\kbar\sqrt D_{cl}}{\sqrt \pi}\, \theta (t-4t_E)\,
\tau_\Phi^{3/2}h\left(\frac{t-4t_E}{\tau_\Phi}\right) \, ,
\label{resultTP}
\end{equation}
where the function $h(x)$ is
\begin{equation}
h(x) = \int_0^x dy_1 \int_0^{\sqrt{y_1}} dy_2\,\, e^{-y_2^2}\, .
\label{H}
\end{equation}
Notice that here $t_E$ is a function of $\Phi$, i.e.
$t_E(\Phi)=\lambda(\Phi)^{-1} \ln \sqrt{K/\kbar}$\,. For
simplicity, we neglected the fluctuations of $\lambda(\Phi)$, i.e.
we put $\lambda_2=0\, $. In the region $t-4t_E\gg \tau_\Phi\,$,
the one-loop correction is exponentially suppressed. One is
required, thus, to consider the higher-loop corrections. The
two--loop correction (Fig.~\ref{fig2loop}) gives the leading weak
dynamical localization correction.

\subsection{Two-loop correction}
\label{twoloop}

In principle the technique developed at the one-loop level can be
employed to treat the two-loop case. However, this is technically
quite involved and is not discussed here. To read out the
frequency-dependent diffusion coefficient in an economical way,
let us renormalize the standard results of the weak
localization\cite{Efetov97} with an appropriate $t_E$-dependent
factor [this procedure indeed is transparent at the one-loop
level, Eq.~(\ref{DDclKW})]. The renormalization factor for the
two-loop geometries (see Fig.~\ref{fig2loop}) were calculated in
Ref.~\onlinecite{TL03} in the context of the Lorentz gas. Adopting
the analogy between the Lorentz gas model and the QKR, verified
above on the one--loop level, one finds for the two--loop
frequency--dependent correction to the diffusion coefficient
\begin{equation}
\delta D (\omega) =-2\kbar^2 D_{cl} \, \Gamma_3(\omega)\!
 \left[\int \!\!\frac{d \varphi}{2\pi\,
 \left(-i\omega+D_{cl}\varphi^2\right)}\right]^2 \, ,
                                                       \label{DDclKW2loop}
\end{equation}
where \cite{TL03,VL03}
\begin{equation}
\Gamma_3 (\omega) =\exp \left(6i\omega t_E -\frac{9\omega ^2
\lambda _2 t_E}{\lambda ^2}\right) \, . \label{Gamma3}
\end{equation}
As a result,  the leading correction to the momentum dispersion in
the case  of broken  TRS is given by:
\begin{equation}
\delta\langle l^2(t)\rangle = 2D_{cl}t - \frac{1}{4}\,\kbar^2\,
\theta (t- 6t_E) \left(t- 6t_E\right)^{2} \, .
                                           \label{result2loop}
\end{equation}
Again we ignore $\lambda_2$ for simplicity. To develop the
two-loop geometry, a minimal quantum wave packet must take time
$t_E$ to expand into some macroscopic size, and vice versa. Within
the logarithmic accuracy  $t_E$ appearing here is the same as that
in the one-loop correction. Therefore, the duration for a full
travel through the Lyapunov region remains the same as the
one-loop case, namely $2t_E$ time. The two-loop geometry involves
three successive visits. In the diagrammatical language, each leg
of the $6$-leg Hikami box (see Fig.~\ref{fig2loop} b) lasts time
$t_E$. Thus ,the weak localization correction, given by this
diagram, is delayed by $6t_E$\, \cite{TL03}.

\section{Observations of Classical-to-Quantum Crossover in Realistic Driven
Systems} \label{EhrDri}

In this section, we discuss some possibilities for experimental
observations  of the predicted $t_E$-dependence of the
classical-to-quantum crossover. The quantity to be measured is the
dispersion function $\delta\langle l^2(t)\rangle$.

\subsection{Energy growth in cold atomic gases}
\label{MDCAG}

In the 90's unprecedented degree of control reached in experiments
with ultra-cold atomic gases \cite{Chu} allowed to investigate
various fundamental quantum phenomena. The advent of laser cooled
atomic gases and standing wave optical pulses
\cite{Raizen95,Raizen99,Ammann} has opened the door to study
quantum chaos experimentally. In an insightful paper,\cite{GSZ}
Graham, Schlautmann, and Zoller pointed out that atom optics may
serve as a testing ground for quantum chaos. Shortly later, the
idea came into realization with sodium atoms being cooled and
trapped using the magneto-optical trap, subjected to a
phase-modulated standing wave \cite{Raizen94}. Later on, a
realization of the QKR in atom optics was accomplished with the
phase-modulated standing wave replaced by a pulsed standing wave
\cite{Raizen95}.

In an atom-optical experiment, typically $10^6$ sodium or cesium
atoms are trapped and cooled down to $10\mu {\rm K}$ using the
conventional magneto-optical trap. After turning off the trapping
fields, two linearly polarized, counter-propagated optical beams
with the frequency $\omega_L$ are switched on, creating a
spatially periodic potential: $V_0\cos (2k_Lx)$\,. Here
$k_L=\omega_L/c$ is the laser wave number. Such optical lattice is
controlled by the acousto--optical modulator as a pulse sequence
with a profile $f(t)$\,. The pulse length $\tau_p$ may be much
smaller than the period $T$\,. The atomic cloud, exposed to this
pulsed optical lattice thereby, experiences a series of kicks. The
evolution of the atomic momenta distribution is monitored after a
certain number of kicks.

In experiments, the laser detuning $\Delta_L\equiv
\omega_L-\omega_0$ from the resonance frequency $\omega_0$ is
large compared to the excited-state decay rate. The dipole force
due to Stark effect leads to the spatially--dependent shift of
atomic levels. This results in an effective periodic potential
imposed on the atomic cloud. One may model the center-of-mass
motion of  atoms with the single-particle time--dependent
Hamiltonian as
\begin{equation}
{\hat H}=\frac{{\hat p}^2}{2m}+
V_0 \cos \left(2k_L {\hat x}\right)\sum_{n=0}^N f(t-nT),
\label{HamiltonianCAG}
\end{equation}
where $m$ is the atomic mass, and $k_L$ is the laser wave number.
The effective potential, $V_0$ is determined by the maximum Rabi
frequency $\Omega$ as $V_0 =\hbar\Omega^2/8\Delta _L$.  The
position ${\hat x}$, and the momentum ${\hat p}$ operators are
canonically conjugated: $[{\hat x}, {\hat p}]=i\hbar$\,. The
effective kicking strength $K$, and the Planck constant $\kbar$
may be expressed then as:
\begin{equation}
K =8\omega_r T^2 \tau_p V_0/\hbar\,, \quad \kbar =8\,\omega_r T
\label{Kkbar}
\end{equation}
with the recoil frequency $\omega_r =\hbar k_L^2/2m$\,. With the
rescaling: $x\rightarrow \theta\equiv 2k_L x$,\, $p\rightarrow
l\equiv (\kbar/2\hbar k_L) p$,\, $t\rightarrow t/T$,\,
$f(t)\rightarrow f(t)/\tau_p$,\, $H\rightarrow (\kbar T/\hbar)
H$\,, one casts the Hamiltonian into Eq.~(\ref{Hamiltonian}) in
the limit $\tau_p/T\rightarrow 0$ (with $K$ fixed)
\cite{footnoteKP}.

So far many experimental efforts \cite{Raizen95,Ammann} have been
focused on the parameter range  where $\kbar\gtrsim 2$, and thus
$t_E \approx 1$. The dynamical localization has been observed as
the saturation of the time-dependent momentum distribution width
(i.e. energy absorption). To extract accurately the
$t_E$-dependent crossover
 one needs a large separation between the relevant time
scales: $1    <      t_E    <      t_L$\,. This requires to
decrease $\kbar$ down to $ 0.1-1 $\,, which we hope to be soon
within the reach for  cold atomic gases experiments
\cite{Raizen00,kbarexp}.

\subsection{Charge fluctuations on Josephson grains}
\label{CFJG}

The experimental realization of the QKR may be also feasible in
experiments involving small nonequilibrium superconducting
systems. One example is a small superconductive dot in contact
with a bulk superconductor through two Josephson junctions
\cite{TKL04a}. The bare Josephson coupling, $E_{J0}$, is modulated
via the external magnetic flux threading the SQUID loop
\cite{CDH98,Haviland1,Haviland2}: $E_J=E_{J0}|\cos \left(\pi
BA_{loop}/\Phi_0\right)|$,
where $A_{loop}$ is the  area of the SQUID loop, and $\Phi_0\equiv
h/2e$ is the superconducting flux quantum. If  $B$ is modulated in
a meander way, with the pulse length much smaller than the period
$T$, the system may be modelled by the Hamiltonian:
\begin{equation}
{\hat H}=\frac{\left({\hat Q}-CV_g\right)^2}{2C} - {\bar E_J} \cos
\theta \sum_n T\delta (t-nT) \, . \label{JGHamiltonian}
\end{equation}
Here $\hat \theta\, ,$ and $ {\hat Q}$ are the relative phase of
the superconducting order parameter on the grain, and its charge,
correspondingly. They are canonically conjugated: $[{\hat
\theta,\, {\hat Q}}]=2e\, i$. In Eq.~(\ref{JGHamiltonian})  $C$ is
the capacitance, $V_g$ is the gate voltage, and ${\bar E_J}$ is
the time-average Josephson coupling. Making  change of the
variables: ${\hat Q}\rightarrow {\hat l} \equiv \hbar {\hat Q}/2e$
and rescaling the relevant quantities as $t\rightarrow
t/T$\,,${\hat H}\rightarrow 8E_cT^2\, {\hat H}/\hbar^2$, and
$CV_g\rightarrow v_g\equiv \hbar CV_g/2e$ ($E_c=e^2/2C$), we cast
the Hamiltonian above into QKR:
\begin{equation}
{\hat H}=\frac{1}{2}\left({\hat l}-v_g\right)^2 -K\cos \theta
\sum_n \delta (t-n) \, . \label{KRJGHamiltonian}
\end{equation}
Note that the sign difference in the kicking term is immaterial.
\begin{figure}[h]
  \centerline{\epsfxsize=3in\epsfbox{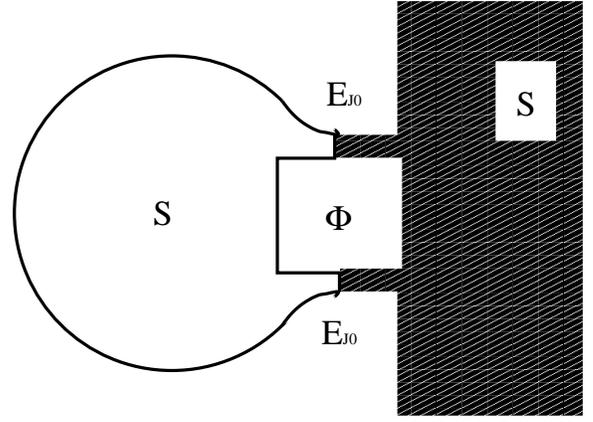}}
\caption{ The scheme of the driven Josephson grain: a
superconducting dot (left) is coupled to  a bulk superconductor
(right) via a SQUID loop (middle). The flux $\Phi$ piercing the
SQUID loop is time-dependent. It effectively modifies the bare
Josephson coupling $E_{J0}$ of the two junctions. }
\label{drivenJJfig}
\end{figure}
\noindent Here the effective Planck's constant and kicking
strength are
\begin{equation}
\kbar =8E_c T/\hbar\, , \quad K= 8E_c{\bar E}_J T^2/\hbar^2 \, ,
\label{hbar}
\end{equation}
respectively.

The charge fluctuations are described by the charge dispersion:
$\sim \delta\langle l^2 (t)\rangle $\,. For sufficiently large $K$
and $t< 4t_E$  it is expected to increase linearly in time.
At $t \gtrsim 4 t_E$, it should deviate from the linearity
\cite{TKL04}. At longer time, $t \gg 4 t_E$ the $t^{3/2}$
power--law correction develops following  the conventional weak
localization theory \cite{Basko,Altland93,AZ96}.
This signals the onset of the localization phenomena. Eventually
at $t\sim D_{cl}/\kbar^2\gg t_E$ the charge fluctuations saturate
and do not grow any more upon further kicking.

\subsection{Charge fluctuations in superconducting nanocircuits}
\label{CFSN}

Recent work \cite{MRBF04} suggested another kind of
time--modulated small  superconducting system. It is proposed that
a mechanically driven superconducting single electron transistor
(SSET) may serve as a realization of the QKR. The system is based
on a Cooper pair shuttle -- a small superconducting island,
periodically travelling between two macroscopic superconducting
leads with the phases $\phi_L$ and $\phi_R$, respectively. Twice
during each period, $T$\,, the shuttle meets one of the leads,
experiencing a sudden Josephson coupling. The average coupling
energy $\bar E_J$ is assumed to be much larger than the charging
energy $E_c$. If the two leads are far enough from each other, the
island never couples to both leads simultaneously. If the
switching time is short, the time-dependent Josephson coupling may
be mimicked by delta-pulses. Therefore, one may model the system
with the following Hamiltonian:
\begin{eqnarray}
{\hat H}           &               =           & -4E_c
\frac{\partial^2}{\partial \theta^2} - {\bar
E_J} \sum_n [ \cos \theta\,\, T\delta (t-2nT) \nonumber\\
           &                          &                + \cos \left(\theta+\Phi\right)\,\,T \delta
(t-(2n+1)T)]. \label{SNHamiltonian}
\end{eqnarray}
\begin{figure}[h]
  \centerline{\epsfxsize=3in\epsfbox{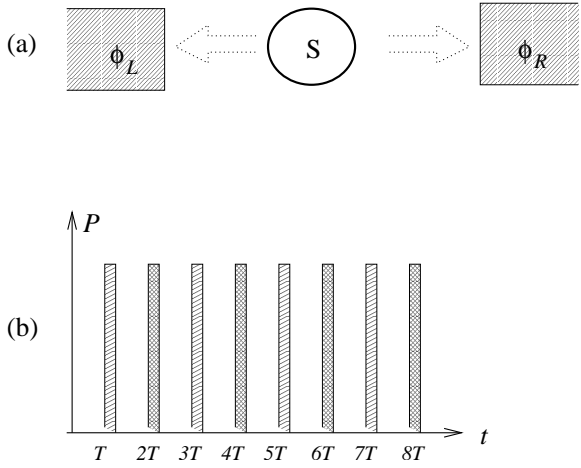}}
\caption{ A superconducting shuttle periodically travels between
two superconducting leads with the phase $\phi_L$ and $\phi_R$\,,
respectively (a). At every other period $2T$\,, the shuttle
experiences a sudden Josephson coupling with the left (right)
lead. } \label{shuttle}
\end{figure}
\noindent Here $\theta$ is the relative phase of the
superconducting island with respect to the right lead. Remarkably,
the phase difference across the two superconducting leads,
$\Phi\equiv \phi_R-\phi_L$ breaks the ``time-reversal'' symmetry,
Eq.~(\ref{TimSym}). The effective Planck  constant and kicking
strength remain the same as in Eq.~(\ref{hbar}). With the same
rescaling as in Sec.~\ref{CFJG}, Eq.~(\ref{SNHamiltonian}) is
rewritten as Eq.~(\ref{SNHamiltonianrescale}).

The classical-to-quantum crossover is reflected in the
nonequilibrium charge fluctuations of the superconducting island.
In the case of $\Phi =0$, the situation is the same as
Sec.~\ref{CFJG}. In the presence of a small phase bias across the
two superconducting leads, i.e., $\left|\Phi \right| \lesssim
\kbar/K$\,, the charge fluctuations are described by
Eq.~(\ref{resultTP}). For larger phase bias: $\left|\Phi
\right|\gtrsim \kbar/K$\,, the charge fluctuations are given by
Eq.~(\ref{result2loop}).

\section{Weak dynamical localization in presence of noise}
\label{noise}

Due to the quantum interference nature of the weak dynamical
localization, the effect may be strongly sensitive to noise.
Indeed, noise effects are known to be of  importance for the
dynamical localization \cite{Raizen00,Cohen91,Milner00,Raizen98}.
So far both experimental and theoretical studies have been
primarily focused on $t_E\lesssim 1$. Below we shall consider how
an external noise affects classical-to-quantum crossover at time
scales $t\gtrsim t_E$. To this end, we will employ the technique
of Sec.~\ref{WDLKR} to investigate a noise--sensitivity of the
weak dynamical localization  Cooperon and Hikami box.

\subsection{Phase noises}
\label{phasenoise}

In proposed experiments involving small superconducting systems,
the phase may fluctuate due to the influence of the dissipative
measurement circuit. For the analytical treatment we focus on the
particular kind of noises -- the Gaussian phase noise. That is,
the phase is assumed to fluctuate randomly  in time:
$\theta_n\rightarrow \theta_n+\zeta _n$\,. Here $\zeta_n\equiv
\zeta_t$ is the noise that is assumed to be uncorrelated at
different kicks $\langle \zeta_n \zeta_{n'}\rangle =0$ for $n\neq
n'$. At a given kick the random phase $\zeta$ is supposed to be
drawn from some {\em periodic}   distribution function
$P(\zeta)=(2\pi)^{-1} \sum_m P_m e^{im\zeta}$. For simplicity we
shall assume that
\begin{equation}
P_m  =  e^{-\sigma_\varphi m^2} \, , \label{phasedistribution}
\end{equation}
where $\sigma_\varphi$ characterizes the strength of the noise.

\subsubsection{Strong phase noise}
\label{largenoise}

The strong noise limit is characterized by
$\sigma_\varphi\rightarrow \infty$, and thus $P(\zeta)=1/2\pi$
being uniform distribution on $\zeta\in [0,2\pi]$\,. We show first
that the classical diffusion (without localization) is restored in
this limit.  The quantum density-density correlator satisfies (cf.
Eq.~(\ref{QDiff})):
\begin{eqnarray}
           &                          &               {\cal D}_{\zeta}(l_+,l_-; l_+',l_-')=
e^{\frac{i\left(l_+^2-l_-^2\right) }{2\kbar}} \, \delta_{l_+,\,
l_+'}\delta_{l_-,\, l_-'}
                                               \label{correlatornoise}\\
           &                          &                + e^{i\omega} \sum_{l_+'',\, l_-''} \overline{\langle
l_+|{\hat U}|l_+'' \rangle
\langle l_-|{\hat U}|l_-'' \rangle ^* }\, {\cal D}_\zeta
(l_+'',l_-''; l_+',l_-';\omega)\, , \nonumber
\end{eqnarray}
where the long bar stands for the average over the phase noise
with respect to the uniform distribution, i.e.,
\begin{eqnarray}
           &                          &               \overline{\langle l_+|{\hat U}|l_+'' \rangle \langle
l_-|{\hat U}|l_-'' \rangle ^*}
 \label{selfenergynoiseaverage}\\
           &               \equiv           &                \int\frac{d\zeta}{2\pi}
\frac{d\theta_+}{2\pi}\frac{d\theta_-}{2\pi} \exp
\left\{\frac{iK}{\kbar}\left[\cos\left(\theta_++\zeta\right)-\cos\left(\theta_-+
\zeta\right)\right]\right\}
\nonumber\\
           &                          &               \times \exp
\left[\frac{i\left(l_+^2-l_-^2\right)}{2\kbar}+
\frac{i\theta_+}{\kbar} \left(l_+-l_+'' \right)
-\frac{i\theta_-}{\kbar}
\left(l_--l_-'' \right) \right] \nonumber\\
           &               =           &                \exp
\left[\frac{i\left(l_++l_-\right)\left(l_+-l_-\right)}{2\kbar}\right]\,
f\left(l_+-l_+''\right) \delta_{l_+-l_-,\, l_+''-l_-''} \,
.\nonumber
\end{eqnarray}
In the last line $f(l)$ is defined as
\begin{equation}
f(l)\equiv\int \frac{d\varphi}{2\pi}\, J_{2l}\!
\left(\frac{2K}{\kbar}\sin\frac{\kbar\varphi}{2}\right) \, .
\label{fn}
\end{equation}

With the definition: $l\equiv (l_++l_-)/2$, $l'\equiv
(l_+'+l_-')/2$, and $\Delta l\equiv l_+-l_-$\,, the solution of
Eq.~(\ref{correlatornoise}) may be formally written as
\begin{eqnarray}
           &                          &               {\cal D}_{\zeta}(l,l';\Delta l)
\label{solutionnoise}\\
           &               =           &               \exp\left[{\frac{il\Delta l}{2\kbar}}\right] \,
\delta_{l,\, l'}+ \sum_{n=1}^{\infty} \, e^{i\omega
n}\sum_{l_1,\cdots ,l_{n-1}} \exp\left[\frac{i\Delta
l}{2\kbar}\sum_{k=1}^n l_k\right] \nonumber\\
           &                          &                \times f\left(l-l_1\right)f\left(l_1-l_2\right)\cdots
f\left(l_{n-1}-l'\right) \, , \nonumber
\end{eqnarray}
where $l_n\equiv l'$\,. As a result of the averaging over noises,
${\cal D}_{\zeta}$ has no $\zeta$-dependence. In the large $K$
limit, we expect the solution to be independent of $(l+l')/2$.
Thus, taking the average over $(l+l')/2$, one finds:
\begin{eqnarray}
\left\langle {\cal D}_{\zeta}(l,l') \right\rangle
           &               =           &               \delta_{l,\, l'}+ \sum_{n=1}^{\infty} \, e^{i\omega
n}\sum_{l_1,\cdots ,l_{n-1}}
\label{solutionnoiseaverage}\\
           &                          &                \times f\left(l-l_1\right)f\left(l_1-l_2\right)\cdots
f\left(l_{n-1}-l'\right) \, . \nonumber
\end{eqnarray}
Passing to Fourier representation, we find that it is nothing, but
the classical diffuson, Eq.~(\ref{DiffClassict1}).  It is worth
mentioning that in the strong noise limit the diffusion constant
is {\em exactly} $K^2/4$. This is due to the fact that higher
order corrections \cite{Rec81} as well as quantum renormalization
\cite{She87,DP02} result from long--time correlation effects (cf.
Appendix~\ref{Difffewkick}). The latter is completely destroyed by
the strong noise.

\subsubsection{Weak phase noise limit}
\label{phasenoisesmall}

We turn now to a more interesting case, relevant to the context
discussed here, i.e., $\sigma_\varphi\ll 1$, where the noise only
slightly suppresses the  weak dynamical localization. For
$\langle\zeta^2 \rangle\sim \sigma_\varphi\ll 1$\,, $\cos
(\varphi+\zeta)\approx \cos\varphi-\zeta \sin\varphi$.

Let us concentrate on the effect of the noise on the Cooperon.
Upon every kicking, Cooperon acquires an additional phase:
$K(\zeta_t \cos\theta_+ - \zeta _{t'} \cos\theta_-)/\kbar$\,.
Recall that $t\,, t'$ are counted from the two opposite ends of
the loop trajectory, and $t+t'\equiv T_{t}$ with $T_{t}$ being the
total duration of the loop.  Since $\zeta_t$ and $\zeta_{t'}$
stand for noises at different moments, they are uncorrelated.
Averaging the phase factor over them leads to the exponential
suppression of of every step of the Cooperon ladder:
\begin{eqnarray}
           &                          &                \left\langle \exp{\left(\frac{iK\zeta_t}{\kbar}\,
\cos\theta_+\right) }\, \exp{\left(-\frac{iK\zeta_{t'}}{\kbar}\,
\cos\theta_-\right) } \right\rangle _{\zeta_t,\, \zeta_{t'}}
\nonumber\\
           &               =           &                e^{-\frac{K^2\sigma_\varphi}{2\kbar^2}}\,I_0^2
\left(\frac{K^2\sigma_\varphi}{4\kbar^2}\right) \, .
\label{phasenoiseaverage}
\end{eqnarray}
($I_0(x)$ is the modified Bessel function). This implies that the
Cooperon is suppressed as
\begin{equation}
{\cal C}_0(t)\, \rightarrow {\cal C}_0(t)\, e^{-t/t_\varphi}
\label{CooperonDephase}
\end{equation}
with the dephasing time defined as
$t_\varphi=2\kbar^2/(K^2\sigma_\varphi)$\,.
For $\sigma_\varphi\gtrsim (\kbar/K)^2$\,, $t_\varphi\sim 1$\,,
the Cooperon mode is suppressed completely. It does not mean,
however, that the classical diffusion is restored for such small
$\sigma_\varphi$. Indeed, the  higher order loop corrections, that
include the Diffusons only (cf. Fig.~\ref{fig2loop}) may still
survive such level of the noise and lead to the  dynamical
localization. To verify if this is the case, one needs to study
the effect of the week noise on Hikami box.

Starting from Eqs.~(\ref{notation2}) and (\ref{notation3}), we
observe that $\delta \theta_1$ and $\theta''$ are independently
shifted by the noise, i.e., $\delta \theta_1\, \rightarrow \delta
\theta_1'\equiv \delta \theta_1+\zeta_1\, , \theta''\, \rightarrow
\theta''+\zeta_2$\,. As a result, instead of Eq.~(\ref{Xres}),
${\cal X}$ is given by
\begin{eqnarray}
           &                          &               {\cal X}\left(\delta l_1,\,\delta\theta_1;\,\delta
l_2,\,\delta\theta_2\right)
\label{Xresnoise}\\
           &               \rightarrow           &                \exp \left(\frac{iK\,\zeta\,
\delta\theta_1'\, \delta\theta_2 }{\kbar}\, \sin \theta''
\right)\, {\cal X}\left(\delta l_1,\,\delta\theta_1';\,\delta
l_2,\,\delta\theta_2\right) \, .
 \nonumber
\end{eqnarray}
Upon  averaging over $\zeta$, Eq.~(\ref{Xresnoise}) leads to the
the exponential suppression of the minimal quantum wave packet
(Hikami box) as
\begin{equation}
\exp\left[-\left(\frac{K\,
\delta\theta_1'\delta\theta_2}{2\kbar}\right)^2\, \sigma_\varphi
\right] {\cal X}\left(\delta l_1,\,\delta\theta_1';\,\delta
l_2,\,\delta\theta_2\right) \, . \label{Hikamisuppression}
\end{equation}
From here we see that the weak phase noises ($\sigma_\varphi \ll
1$) does {\it not} substantially affect Hikami box. Indeed, the
interaction vertex is significantly suppressed only at
$\sigma_\varphi\sim 1 $\,. This means that the effect of phase
noises on Hikami box may be ignored compared to the dephasing of
Cooperon.

As a result, the intermediate intensity noise $(\kbar/K)^2\ll
\sigma_\varphi\ll 1$ acts, to much extent, as a TRS breaking
perturbation. It suppresses the Cooperon corrections, leaving the
Diffuson ones (the simplest being the two loop one
Fig.~\ref{fig2loop}) intact.

Above we find that the Diffuson-Cooperon coupling--minimal wave
packet, is suppressed by large enough phase noises. This picture
is naturally expected to be applicable for higher order
interaction vertex also. The latter is essentially responsible for
the onset of the weak dynamical localization in Diffuson-only
diagrams. This picture may be considered to be the precursor of
the restoration of diffusion for larger noises.

\subsection{Amplitude noise}
\label{amplitudenoise}

In  cold atoms experiments, the optical pulse power may fluctuate
with time and thus the series is not perfectly periodic. This then
leads to the noise in the kicking amplitude, i.e. the stochastic
parameter $K$ is replaced by $K+\eta_n$, where $\eta_n=\eta_t$ is
a random amplitude fluctuation. The effects on dynamical
localization of such kind of noise have been under intensified
experimental investigations \cite{Raizen00,Milner00,Raizen98}. A
central issue addressed is, whether the dynamical localization is
destroyed completely by the noise. We  concentrate here on a weak
noise limit.

To simplify analytical estimations below let us assume that
$\eta_n$ is the white--noise, Gaussian noise, namely $\langle
\eta_n \eta_{n'}\rangle =\sigma_K\delta_{nn'}$.  Here $\sigma_{\rm
K}$ characterizes the strength of the noise. Below we consider the
limiting case of $\sigma_{\rm K}\ll 1$, where the amplitude noise
only slightly suppresses the  weak dynamical localization.

For the Diffuson the one--step quantum propagator acquires an
additional phase as $\eta_t (\cos\theta_+-\cos\theta_- )$.
Here $\theta_+$\,, $\theta_-$ stand for the phases of
retarded/advanced Green's functions at the kicks. Passing to the
semiclassical limit, i.e., $|\theta_+-\theta_-|\ll 1$ and
averaging over the noise, we find that the Diffuson is affected
via  renormalization of  $D_{cl}$ by a small correction $\delta
D_{cl} \sim \sigma_{\rm K}\ll D_{cl}$ (A detailed study in this
direction was presented in Ref.~\onlinecite{Raizen99}.). This
correction is not responsible for the destruction of the dynamical
localization. It is the effect of  the noise on the Cooperon and
Hikami box that  eventually may lead to the restoration of the
classical diffusion.

In the case of the amplitude noise the effect on the Cooperon is
fully analogous to the case of the phase noise, i.e. dephasing.
Upon every kicking, the Cooperon acquires an additional phase:
$(\eta_t \cos\theta_+ - \eta _{t'} \cos\theta_-)/\kbar$ with $t\,,
t'$ counted from the opposite ends respectively, such that
$t+t'\equiv T_{t}$ with $T_{t}$ being the total duration of the
loop.
Upon averaging over the independent noises $\eta_t$ and
$\eta_{t'}$, this phase factor leads to the exponential
suppression of a single step of the Cooperon as:
\begin{eqnarray}
           &                          &                \left\langle \exp{\left(\frac{i\eta_t}{\kbar}\,
\cos\theta_+\right) }\, \exp{\left(-\frac{i\eta_{t'}}{\kbar}\,
\cos\theta_-\right) } \right\rangle _{\eta_t,\, \eta_{t'}}
\nonumber\\
           &               =           &                \exp \left[-\frac{\sigma_{\rm K}}{2\kbar^2}\right] \,
I_0^2\left(\frac{\sigma_{\rm K}}{4\kbar^2}\right)\, .
\label{amplitudenoiseaverage}
\end{eqnarray}
This implies that the Cooperon is  dephased as
Eq.~(\ref{CooperonDephase}) with the dephasing time being
$\tau_{\rm K} =2\kbar^2/\sigma_{\rm K}$\,.
For $\sigma_{\rm K}\gtrsim \kbar^2$\,, $\tau_{\rm K}\sim 1 $\,,
the Cooperon mode is suppressed completely thus only higher order
terms (Diffusons only) may be responsible for the  dynamical
localization. It remains open to estimate the noise amplitude that
destroys Diffuson-only weak dynamical localization.

\subsection{Effects of finite dephasing time}
\label{dephasing}

We saw above that every realization of the QKR may involve various
noises in realistic experimental environments. As a result, there
exist various dephasing mechanisms. The effective dephasing rate
is the sum over all dephasing rates, namely
\begin{equation}
\frac{1}{\tau_\varphi}=\sum_k \frac{1}{\tau_k} \, .
\label{totaldephasing}
\end{equation}

To simplify qualitative discussions, in this part we focus on
effects arising from weak dephsing such that $4t_E \ll
\tau_\varphi$. First we show that the effective Ehrenfest time is
shortened. In fact, the additional weak noises enhance the rate of
angular deviation spread according to:
\begin{equation}
\frac{d}{dt} \, \delta \theta^2 (t) = 2 \lambda \, \delta \theta^2
(t) + \frac{1}{\tau_\varphi} \, . \label{deviationevolution}
\end{equation}
Recall that the first term on the right hand side results from the
Lyapunov instability.  The solution is easily found, i.e., $\delta
\theta^2 (t) = [\delta \theta_{1,2} + (2\lambda
\tau_\varphi)^{-1}] \, e^{\lambda t} - (2\lambda
\tau_\varphi)^{-1}$. Here $\delta \theta_{1,2} $ are the initial
angular deviation of Diffuson/Cooperon, respectively. At $t^{{\rm
D}*}$ (or $t^{{\rm C}*}$), $\delta \theta \sim 1 $. The solution
then gives
\begin{equation}
t^{{\rm D, C}*} \approx \frac{1}{\lambda} \left| \ln \frac{1}{
\sqrt {\delta \theta_{1,2}^2 + (2\lambda
\tau_\varphi)^{-1}}}\right| \, . \label{te}
\end{equation}
Thus, the effective duration of a full travel through the Lyapunov
region is found to be
\begin{eqnarray}
    &    &     t^{{\rm D}*} + t^{{\rm C}*}   = \nonumber\\
    &    \frac{1}{\lambda}     &     \left|
\ln \frac{1}{ \sqrt {\left(\delta \theta_1\, \delta \theta_2
\right)^2 + \left(\delta \theta_1^2 + \delta \theta_2^2 \right)\,
(2\lambda \tau_\varphi)^{-1} + (2\lambda
\tau_\varphi)^{-2}}}\right|
\nonumber\\
   &   \leq   &   \frac{1}{\lambda}
\left | \ln \frac{1}{ \left | \delta \theta_{1}\, \delta
\theta_{2}
\right | + (2\lambda \tau_\varphi)^{-1}} \right| \nonumber\\
   &   \sim   &
 \frac{2}{\lambda} \left | \ln
\frac{1}{\sqrt \frac{\kbar}{K}+ \left (2\lambda \tau_\varphi
\right)^{-1}} \right | = 2 t_E^* \, . \label{teEffective}
\end{eqnarray}

The effective Ehrenfest time $t_E^*   <    t_E$. With $t_E^*$
substituted into the renormalization factors ${\cal W}_{\rm C}$
and ${\cal W}_{\rm D}$, and taking into account the dephasing of
diffusive Cooperon, the one-loop correction, Eq.~(\ref{DDclKW}),
is modified as
\begin{eqnarray}
\delta D(\omega)            &               =           & -{\kbar
D_{cl}\over \pi}\, {\cal W}_{\rm D}\left(2\omega\right){\cal
W}_{\rm C}\left(2\omega\right)\nonumber\\
           &                          &               \times
 \int \!\! \frac{d
 \varphi}{-i\omega+D_{cl}\varphi^2+\frac{1}{\tau_\varphi}} \, .
                                                       \label{DDclTP}
\end{eqnarray}
Taking the Fourier transform with respect to $\omega$, one find
the momentum dispersion to be
\begin{eqnarray}
     &        &      \delta\langle l^2(t)\rangle
     \label{resultdephasing}\\
     &    =    &      2D_{cl}t - \frac{8\kbar \sqrt{D_{cl}}}{3\sqrt \pi}\,\theta (t-
4t_E^*) \left(t- 4t_E^*\right)^{3/2} \,
e^{-\frac{t-4t_E^*}{\tau_\varphi}} \, .
                                           \nonumber
\end{eqnarray}
Again $\lambda_2 = 0$. According to Eq.~(\ref{resultdephasing})
the quantum correction is suppressed at $t \gtrsim 4t_E^* +
\tau_\varphi $.

One should keep in mind that in realistic experiments, like
atom-optical ones the Hamiltonian of QKR, Eq.~(\ref{Hamiltonian})
may be an oversimplification. Therefore, some restriction on the
validity of the present result will be imposed. For example, the
cold atoms experiment involves collisions between the atoms, which
lead to another dephasing mechanism. Let us  estimate the
corresponding collision dephasing time, $\tau_s$. The two-particle
scattering mean free path is known to be $l_s\approx 1/(na^{2})$,
where $n$ is the atomic concentration and $a$ is the $s$--wave
scattering length. The corresponding dimensionless scattering time
is $\tau_s = l_s/(Tv)$, where $v$ is a typical atomic velocity
that may be estimated as $v\approx \hbar k_L \, |l|/ \kbar m
\approx \hbar k_L\, \sqrt{D_{cl}\tau_s}/\kbar m\,$. This leads to
the self-consistent estimate of the dimensionless dephasing time:
\begin{equation}
\tau_s \approx (l_s k_L/K)^{2/3}. \label{TPCAG}
\end{equation}
Once again to observe the classical--to--quantum crossover the
inequality  $4t_E \lesssim \tau_s$\, should be valid.

\section{Conclusions}
\label{CON}

In this paper, we developed an analytical theory to incorporate
systematically the Ehrenfest time, $t_E$, into the  weak dynamical
localization. We map the loop expansion, central to the theory of
weak dynamical localization onto the interference of paths in the
configuration space with a certain loop geometry. To propagate
along such paths a time longer than $2mt_E\, (m=2,3,\cdots)$
(where $m$ is determined by a specific loop geometry) is needed.
We establish thus that the onset of the dynamical localization is
delayed by the multiples of $2t_E$. In particular, for QKR, the
delay is $4t_E$, while for  systems with broken time-reversal
symmetry it is $6t_E$. At shorter times, quantum corrections to
the linear dispersion (classical diffusion) {\it do} exist.
However, they results only  in the renormalization of the
frequency-independent diffusion coefficient, and thus are {\it
not} responsible for the onset of the dynamical localization.

Our quantitative predictions are based on the loop expansion and
are essentially perturbative. They are closely analogous to the
weak Anderson localization in ballistic systems. They may be
considered as a further support for  the long-standing conjecture
of Ref. \onlinecite{FGP82} about the similarity between the
dynamical localization and Anderson localization for irrational
$\kbar/4\pi$ \, \cite{Izrailev90}. The perturbative corrections
are responsible only for the early evolution of the dispersion
function, i.e., the suppression of the classical diffusion. At
longer time $t\gtrsim t_L$, the strong dynamical localization is
expected to develop, as supported by arguments based on
supersymmetric diffusive $\sigma$ model \cite{AZ96}. However, it
still remains a  challenge to prove this conjecture in the
presence of finite $t_E$.

Although our analytical treatment is developed  for the QKR, we
believe  that it may be extended to more general quantum driven
systems. The essential requirements imposed on the dynamics of an
underlying classical system are: (i) area--preserving; and (ii)
the existence of classical stochastic diffusion (subject to
certain symmetry). Provided (i) and (ii)  are satisfied, we expect
the  functional form of the weak localization corrections to be
the same as derived above. The specific of chaotic motion enters
through modifying the classical quantities such as diffusion
coefficient, Lyapunov exponent and its fluctuations -- such
problem is purely classical and, in general, may be resolved (say
numerically).

Technically, the most important part of this work is the
derivation of the one-loop vertex (Hikami box) {\it without}
introduction of a regularization. This allows to analyze
accurately the minimal quantum wave packet. As a result, the
Ehrenfest time is quantitatively defined (with the logarithmic
accuracy) as the time needed to expand an initial minimal wave
packet up to a macroscopic size \cite{LO68,BZ78}. Alternatively,
the minimal wave packet may be analyzed within the Moyal formalism
\cite{Zirnbauer99,SET03}. A study along this line has been
reported recently in the context of ballistic supersymmetric
$\sigma$-model \cite{MA04}. This supports the conjecture of
Ref.~\onlinecite{AL96} that the Ehrenfest time should not depend
on the regularization, since the later is only intended to mimic
the effect of quantum diffraction. Indeed, in accord with
Ref.~\onlinecite{AL96}, our current results may be fully
reproduced by introducing a proper regularization
\cite{Zirnbauer99} to the supersymmetric $\sigma$ model developed
for QKR \cite{AZ96}.

The quantitative predictions made for the $t_E$-dependent
classical-to-quantum crossover in QKR may be suitable for
experimental verifications in various contexts. In particular,
already existing experiments on the dynamical localization in the
energy growth of ultra-cold atomic gases have greatly  contributed
to understanding of this crossover. Our quantitative predictions
are expected to be accurate  in the asymptotic regime  $\kbar < 1
<       K$. For their quantitative verification, it is thus highly
desirable  to decrease $\kbar$ down to $0.1-1$. We also pointed
out that  some periodically driven mesoscopic superconducting
structures may be suitable for realizations of the QKR. The
dynamical localization and classical--to--quantum crossover in
these systems are observable by monitoring charge fluctuations of
the superconducting island. It is important to mention that, all
realistic experiments  introduce noise and thus a finite dephasing
time $\tau_\varphi$\,. We have shown here that for an observation
of the  $t_E$-dependent crossover, the condition $4t_E \lesssim
\tau_\varphi$ must be satisfied.

\begin{acknowledgments}
We have greatly benefited from discussions with A. ~Altland, D.
~Basko, S. ~Fishman, J. ~Liu and C. ~Zhang. We  thank L. Glazman
for pointing Ref.~\onlinecite{IGSGZ02} to us. We are grateful to
Abdus Salam International Center for Theoretical Physics, where
part of this work was done, for its hospitality.  C.~T. and A.~L.
are supported by NSF Grant No. DMR-0120702 and DMR-0439026. A.~K.
is A.~P.~Sloan fellow and supported by the NSF grant No.
DMR--0405212.
\end{acknowledgments}

\begin{appendix}

\section{Finite time correlation effects on the self-energy}
\label{Difffewkick}

In this Appendix we study the higher order time correlation
effects, starting from the exact quantum density-density
correlator, Eq.~(\ref{DDDefSig}). In particular, we clarify that
in the semiclassical limit, the higher order corrections, namely
Eq.~(\ref{Dcl}), (see Ref.~\onlinecite{Rec81}) to the diffusion
constant, i.e., $K^2/4$  are found.

For an arbitrary $\tau_c> 1$  Eq.~(\ref{DiffPropoOneKick}) is
replaced  by:
\begin{equation}
{\cal D}_0\left(l,l'\right)= \delta_{l,l'}+\sum_{l''} \Sigma
\left(l,l''\right) {\cal D}_0\left(l'',l'\right) \, .
\label{DDifCoe}
\end{equation}
Here the self-energy, $\Sigma \left(l,l''\right)$ is given by
\begin{eqnarray}
           &                          &                \Sigma \left(l,l''\right)= \sum_{n=1}^{\tau_c}
e^{i\omega n} \langle l|{\hat U}^n |l''\rangle \langle l|
 {\hat U}^n |l''\rangle^*   \label{Dyson} \\
   &    =   &    e^{i\omega} U_{l,l''}U_{l,l''} ^* + \sum_{n=2} ^{\tau_c}
           e^{i\omega n} \sum_{l_{1+},\, l_{1-}} \cdots
\sum_{
  l_{(n-1)+},l_{(n-1)-}} \nonumber\\
    &    &  \times
\prod_{k=1}^{n-1} U_{l_{k+},l_{(k+1)+}}U_{l_{k-},l_{(k+1)-}} ^*
\,. \nonumber
\end{eqnarray}
Note that $l_{0}=l_{0}'\equiv l$, $l_{n}=l_{n}'\equiv l''$, and
$l_{k}  \neq  l_{k}' $ for $0  <  k   <   n$. Here in order to
simplify the notation we denote the matrix elements $\langle
l_{k+}|{\hat U} |l_{(k+1)+}\rangle$ as $U_{l_{k+},l_{(k+1)+}}$,
and similarly for their complex conjugates. These matrix elements
may be written explicitly as
\begin{eqnarray}
U_{l_{k+},l_{(k+1)+}}            &               =           &
\int \frac{d\theta_{(k+1)+}}{2\pi} \exp \bigg
[\frac{il_{(k+1)+}^2}{2\kbar} \nonumber\\
+\frac{iK}{\kbar}   &   \cos  &   \theta_{(k+1)+}
            -\frac{i}{\kbar}\left(l_{k+}-l_{(k+1)+}\right)\theta_{(k+1)+}
\bigg ] \, ,
\nonumber\\
U_{l_{k-},l_{(k+1)-}} ^*            &               =           &
\int \frac{d\theta_{(k+1)-}}{2\pi} \exp \bigg
[-\frac{il_{(k+1)-}^2}{2\kbar}  \label{matrixelement}\\
-\frac{iK}{\kbar}            &    \cos      &     \theta_{(k+1)-}
           +\frac{i}{\kbar}\left(l_{k-}-l_{(k+1)-}\right)\theta_{(k+1)-}
\bigg ] \, . \nonumber
\end{eqnarray}
To proceed further, we introduce the following quantities:
\begin{equation}
m_k=\left(l_{k+}-l_{k-}\right)/\kbar \, , \quad
q_k=\left(\theta_{k+}-\theta_{k-}\right)/\kbar \, . \label{Defmq}
\end{equation}
Then with the substitution of Eq.~(\ref{matrixelement}) we rewrite
$U_{l_{k+},l_{(k+1)+}}U_{l_{k-},l_{(k+1)-}} ^*$ as
\begin{eqnarray}
     &       &      U_{l_{k+},l_{(k+1)+}}U_{l_{k-},l_{(k+1)-}} ^*
     \label{UU} \\
&   = &
   \int\!\!\! \int \frac{d\theta _{(k+1)+}}{2\pi}\frac{d\theta
_{(k+1)-}}{2\pi}\, \exp\bigg \{ i \, m_{k+1}
\frac{l_{(k+1)+}+l_{(k+1)-}}{2} \nonumber\\
    &     &    - \frac{2iK}{\kbar}\,\,
\sin \frac{\kbar q_{k+1}}{2}\sin
\frac{\theta_{(k+1)+}+\theta_{(k+1)-}}{2} \nonumber\\
    &     &    -iq_{k+1}\left[\frac{l_{k+}+l_{k-}}{2}-\frac{l_{(k+1)+}+l_{(k+1)-}}{2}
\right]  \nonumber\\
    &     &    +i \left(m_k-m_{k+1}\right)
\frac{\theta_{(k+1)+}+\theta_{(k+1)-}}{2} \bigg \} \, . \nonumber
\end{eqnarray}
Furthermore, we insert the Fourier transform:
\begin{eqnarray}
    &   &   \exp\left[\frac{2iK}{\kbar} \sin\theta\, \sin \frac{\kbar q}{2} \right]
\nonumber\\
    &  =    &   \sum_n J_{n\, {\rm sgn } \, q }
\left(\frac{2K}{\kbar}\sin \frac{\kbar q}{2}\right)\, e^{in\theta}
\label{BesselFourier}
\end{eqnarray}
into it with ${\rm sgn }$ denoting the sign of $q$. Then
Eq.~(\ref{UU}) is substituted into Eq.~(\ref{Dyson}). With the sum
with respect to $\left(l_{k+}+l_{k-}\right)/2$
($\,\,\,k=0,1,2,\cdots$) and the integral with respect to
$\left(\theta_{k+}+\theta_{k-}\right)/2$ performed, eventually
Eq.~(\ref{Dyson}) is reduced to
\begin{eqnarray}
    &      &    \Sigma(l,l'')   =  e^{i\omega} J_0 \left(\frac{2K}{\kbar}\sin
\kbar
q_0\right)  \label{SigmaQuantum}\\
     &    &    + \sum_{r=2}^{\tau_c}
e^{i\omega r} \sum_{n_1,n_2, \cdots , ,n_r} \sum_{m_1,m_2, \cdots
, m_{r-1}} \sum_{q_0,q_1, \cdots , q_{r-1}} \prod_{k=1}^r
\nonumber\\
     &    &     \times J_{n_k\, {\rm sgn} \, q_{k-1}}\left(\frac{2K}{\kbar}
     \left|\sin \frac{\kbar q_{k-1}}{2}\right|\right)  \nonumber\\
     &    &     \times  \delta_{q_k-q_{k-1}, -m_k}\, \delta_{m_k-m_{k-1},
- n_k \, {\rm sgn}\, q_{k-1}} \, . \nonumber
\end{eqnarray}
Shortly later we will see that the following relations, implied by
the two Kroneck's symbols,
\begin{equation}
q_k=q_{k-1}-m_k  \label{qEq}
\end{equation}
and
\begin{equation}
m_k=m_{k-1}-n_k\, {\rm sgn}\, q_{k-1}  \label{mEq}
\end{equation}
are essential for the derivation of higher order corrections to
the diffusion constant \cite{Rec81}. It is in order to emphasize
that they are exact even at the quantum mechanical level, although
originally found in the classical context \cite{Rec81}.

Note that above $m_0=m_n=0$. So far the discussions above are
formally accurate. We assume now that the correlator, ${\cal D}_0
(l,l')$ does not depend on the center-of-mass, namely ${\cal D}_0
(l,l')={\cal D}_0 (l-l')$. This is supplemented by performing
Fourier transformation for ${\cal D}_0 (l-l')$:
\begin{equation}
{\cal D}_0 \left(l-l'\right)= \int \frac{d\varphi}{2\pi}\, {\cal
D}_0 \left(\varphi\right) \, e^{i\varphi  \left(l-l'\right)}
\label{selfenergyFourier}
\end{equation}
and imposing the boundary condition as (resulting from finite
$\tau_c$ \,\, \cite{Rec81})
\begin{equation}
q_0=q_{r-1}=\varphi \rightarrow 0  \label{qmCon}
\end{equation}
with $\varphi$ denoting the Fourier component. Then Fourier
transforming $\Sigma(l-l')$ leads to
\begin{eqnarray}
   &    &   \Sigma(\varphi)    =   e^{i\omega} J_0 \left(\frac{2K}{\kbar}\sin
\frac{\kbar \varphi}{2}\right)
\label{SigmaQuantumFourier} \\
   &    &    + \sum_{r=2}^{\tau_c}\, e^{i\omega
r}  \sum_{n_1 \neq 0}\sum_{n_2, \cdots , n_r} \sum_{m_1,m_2,
\cdots , m_{r-1}} \sum_{q_1, \cdots , q_{r-2}} \prod_{k=1}^r
\nonumber\\
    &     &    \times  J_{n_1} \left(\frac{2K}{\kbar}\sin \frac{\kbar \varphi}{2}\right)
J_{n_r} \left(\frac{2K}{\kbar}\sin \frac{\kbar \varphi}{2}\right)
\nonumber\\
    &     &    \times   J_{n_k\, {\rm sgn}
q_{k-1} }\left(\frac{2K}{\kbar} \left|\sin \frac{\kbar
q_{k-1}}{2}\right|\right)
\, . \nonumber
\end{eqnarray}
Note that above we suppressed the two Kroneck's symbols to
simplify the expression. One should keep in mind that the sum over
$m$'s, $q$'s and $n$'s are restricted by the two ``motion''
equations, i.e., Eq.~(\ref{qEq}) and (\ref{mEq}).

Let us make the semiclassical approximation, i.e., $|\kbar q_k|\ll
1$ in Eq.~(\ref{SigmaQuantumFourier}) and focus on the limit
$K\varphi\ll 1$. The first term in the self-energy leads to the
diffusion constant as $K^2/4$, as discussed in
Sec.~\ref{DiffApprox}. The second term gives higher order
oscillatory corrections. In fact, up to $(K\varphi)^2$, the sum is
contributed by a particular series of $(q_k,\, m_k) \,
(k=0,1,2,\cdots)$ (so-called Fourier paths) \cite{Rec81} as
$(\varphi,\, 0) \rightarrow \pm (0,\, 1) \cdots \pm (1,\, -1)
\rightarrow (\varphi,\, 0) $. Here $\cdots$ is shorthand of
product of Bessel functions, which is an expansion in powers of
$K^{-1/2}$. Since for $K\varphi\ll 1$, $J_n(K\varphi) \sim
(K\varphi) ^n$, therefore, $n_{1,r} =\pm 1$ is the only
contribution to the order $(K\varphi)^2$. This then leads to
$\Sigma(\varphi)$ as \cite{Rec81}
\begin{eqnarray}
   &   &   \Sigma \left(\varphi\right) = e^{i\omega}J_0(K\varphi) \label{SigPhi}\\
   &    +   &    \sum_{r=2}^{\tau_c}\, e^{i\omega
r}  \sum_{n_1, n_r =\pm 1} \sum_{n_2, \cdots , n_{r-1}}
\sum_{m_1,m_2, \cdots , m_{r-1}} \sum_{q_1, \cdots , q_{r-2}}
\prod_{k=1}^r
\nonumber\\
   &    \times   &    J_{n_1}(K\varphi) J_{n_r}(K\varphi) J_{n_k\, {\rm sgn}
q_{k-1} }\left(2K \left| q_{k-1}\right|\right) \nonumber
\end{eqnarray}
In the limit $\omega\tau_c\ll 1\, , K\varphi\ll 1$, one recovers
the diffuson Eq.~(\ref{DiffAltland}) with the diffusion constant
given by Eq.~(\ref{Dcl}) up to ${\cal O}(1)$ (for $K\gg 1$)
\cite{Rec81}.

Proceeding along this line, we may reproduce Shepelansky's results
for quantum diffusion constant of early evolution
\cite{She87,DP02}. The basic observation is that if number of
kicks is less than $4$, then $q_k=1$. Based on
Eq.~(\ref{SigmaQuantumFourier}), this then implies that the
results of Eq.~(\ref{Dcl}) still holds except that $K$ is replaced
by $2K \sin(\kbar/2)/ \kbar$.

\section{Two relations resulting from TRS}
\label{relation}

In this Appendix we show two exact relations reflecting TRS. In
the first, for any $n     >       0$\,,
\begin{equation}
{\cal D}_0\left(l,\theta;l',\theta';n\right)= {\cal
D}_0\left(l',-\theta';l,-\theta;n\right) \, . \label{DDrelatime}
\end{equation}
{\it Proof.} Use the mathematical deduction. For $n=1$\,,
\begin{eqnarray}
           &                          &                {\cal D}_0\left(l,\theta;l',\theta';1\right) \label{DDrelan1}\\
           &               =           &                \overrightarrow{P}\, \delta \left(l-l'\right)\, \delta
\left(\theta -\theta' -l' \right) \nonumber\\
           &               =           &                \delta \left(l-K\sin (\theta - l)-l'\right)\, \delta
\left(\theta -l - \theta' -l' \right)\, . \nonumber
\end{eqnarray}
On the other hand,
\begin{eqnarray}
           &                          &                {\cal D}_0\left(l',-\theta';l,-\theta ;1\right)
\label{DDrelan12}\\
           &               =           &                \overrightarrow{P}\, \delta \left(l'-l\right)\, \delta
\left(-\theta' +\theta -l \right) \nonumber\\
           &               =           &                \delta \left(l'+K\sin (\theta' + l')-l\right)\, \delta
\left(-
\theta' -l'+ \theta -l \right) \nonumber\\
           &               =           &                \delta \left(l'+K\sin (\theta - l) -l\right) \, \delta
\left( \theta -l - \theta' -l' \right) \, . \nonumber
\end{eqnarray}
Comparing the last two lines of Eqs.~(\ref{DDrelan1}) and
(\ref{DDrelan12}), we immediately see that Eq.~(\ref{DDrelatime})
holds for $n=1$\,. Next we assume that Eq.~(\ref{DDrelatime})
holds for arbitrary $n=k     >       1$\,. Then for $n=k+1$\,, we
obtain:
\begin{eqnarray}
           &                          &                {\cal D}_0\left(l',-\theta';l,-\theta ;k+1\right)
\label{DDrelank}\\
           &               =           &                \overrightarrow{P}\, {\cal
D}_0\left(l',-\theta';l,-\theta ;k\right)
\nonumber\\
           &               =           &                \overrightarrow{P}\, {\cal
D}_0\left(l,\theta;l',\theta' ;k\right)
\nonumber\\
           &               \equiv           &                {\cal D}_0\left(l,\theta ;l',\theta' ;k+1\right)
\, . \nonumber
\end{eqnarray}
Thus, Eq.~(\ref{DDrelatime}) also holds for $n=k+1$. Q. E. D.

Now we turn to show the other relation. That is, if
\begin{equation}
f \left(l,\theta;l',\theta'\right)= f
\left(l',-\theta';l,-\theta\right) \, , \label{condition}
\end{equation}
then
\begin{equation}
\overrightarrow{P}\, f \left(l,\theta;\, l',\theta' \right)= f
\left(l,\theta;\, l',\theta' \right) \, \overleftarrow{P}_T \, .
\label{leftrightPrela}
\end{equation}
{\it Proof.}
\begin{eqnarray}
           &                          &                f \left(l,\theta;\, l',\theta' \right) \,
\overleftarrow{P}_T \nonumber\\
           &               =           &                f \left(l,\theta;\, l'+K\sin (\theta'+l'),\theta'+l'
\right) \nonumber\\
           &               =           &                f \left(l'+K\sin (\theta'+l'),-(\theta'+l');\,
l,-\theta \right) \nonumber\\
           &               =           &                \overrightarrow{P}\, f
\left(l',-\theta';\, l,-\theta \right) \nonumber\\
           &               =           &                \overrightarrow{P}\, f \left(l,\theta;\, l',\theta'
\right) \, , \label{leftrightP}
\end{eqnarray}
where we used the definition, Eq.~(\ref{PFTDef}), in the second
line,  Eq.~(\ref{condition}) in the third and fifth lines, and the
definition, Eq.~(\ref{PFDef}), in the fourth line. Q. E. D.

\section{Derivation of renormalization factor of diffusive Cooperon}
\label{PropagatorLyapunov}

In this Appendix, we  derive the renormalization  factor,
Eq.~(\ref{Gamma}), describing modification of the diffusive
Cooperon due to the propagation through the Lyapunov region. We
first analyze the asymptotic instability of a generic chaotic
trajectory.

\subsection{Asymptotic instability}
\label{asyminstab}

By varying the equations of motion of the classical kicked rotor:
\begin{eqnarray}
\frac{d\theta}{dt}            &               =           &                l, \nonumber\\
\frac{dl}{dt}            &               =           & K
\sin\theta\sum_n \delta\left( t-n \right)
 \label{HE}
\end{eqnarray}
and defining the variables: $z\equiv\ln |\delta\theta|$, $\alpha
\equiv\delta l/\delta\theta$, we obtain:
\begin{eqnarray}
\frac{dz}{dt}            &               =           &                \alpha,\nonumber\\
\frac{d\alpha}{dt}+\alpha^2            &               = &
K \cos \theta \sum_n \delta (t-n), \label{SF}
\end{eqnarray}
which describes the evolution of  separation of two nearby
trajectories along a reference trajectory, initiated from $(l_0\,
,\theta_0)$. From Eq.~(\ref{SF}) we see that $\alpha$ is a fast
changing variable. That is, the dynamics of $\alpha$  introduces
some classical time scale, beyond which the slow changing variable
-- $z$ is independent of initial $\alpha$. Let us further
introduce the variables: $\alpha_n$--denoting $\alpha$ right after
the $n$-th kicking, and $z_n$--denoting $z$ at the $n$-th kicking.
Equivalently, we rewrite Eq.~(\ref{SF}) as
\begin{eqnarray}
z_{n+1}-z_n   &   =    &    \ln (1+\alpha_n) \, , \nonumber\\
\alpha|_{n+1}   &    =    &    \frac{1}{\alpha^{-1} |_{n}+1}
+K\cos\theta_{n+1} \, .
\label{f_between_1}
\end{eqnarray}
For $n \gg 1$, $\langle z_n \rangle = \lambda \, n$, $\langle (z_n
-\lambda \, n )^2 \rangle = 2 \lambda_2 \, n $. Thus, the Lyapunov
exponent $\lambda$ and its fluctuation $\lambda_2$, characterizing
the long time instability are defined as
\begin{eqnarray}
\lambda            &               =           &
\lim_{n\rightarrow\infty} \lambda (n), \quad \lambda_2
=\lim_{n\rightarrow\infty} \lambda_2(n) \, , \nonumber\\
\lambda (n)            &               =           & \frac{1}{n}
\sum_{n'=0}^{n-1} \ln \left(1+\alpha |_{n'}\right)
\, , \label{lamdef}\\
\lambda _2 (n)            &               =           &
\frac{1}{n} \left[ \sum_{n'=0}^{n-1} \ln \left(1+\alpha
|_{n'}\right)-n\lambda \right]^2 \, , \nonumber
\end{eqnarray}
respectively. Note that at finite times, they are
trajectory-dependent. In the limit $n\rightarrow\infty$, the time
average is expected to be equivalent to the average over $\alpha$,
as well as the initial conditions located in the stochastic
region.

The estimation of $\lambda$ and $\lambda_2$ in the limit $K\gg 1$
may be made as follows based on the simple analysis above:
\begin{eqnarray}
\lambda            &               \equiv           &
\left\langle\, \ln|K\cos\theta| \right\rangle =\ln (K/2)\, ;
                                           \label{lambda}\\
\lambda _2            &               \equiv           &
\left\langle\, \ln^2|K\cos\theta| \right\rangle
-\lambda^2=\zeta(3)-\ln^2 2\approx 0.82\, ,
                                                 \nonumber
\end{eqnarray}
where the angular brackets imply uniform averaging over the angle.

\subsection{Renormalization factor}
\label{PropagatorW}

According to the definition of $n_{\rm c}$,
Eq.~(\ref{ncDefinition}) it is the kick number when $z\approx 0$
with the logarithmic accuracy. That is,
\begin{equation}
0=z_0+\sum_{j=0}^{n_{\rm c}-1} \, \ln \left(1+\alpha|_j\right) \,
, \label{zmotion}
\end{equation}
equivalently,
\begin{equation}
n_{\rm c}=-\frac{1}{\lambda (n_{\rm c})}\,
\left\{z_0+\sum_{j=0}^{n_{\rm c}-1} \, \left[\ln
\left(1+\alpha|_j\right)- \lambda (n_{\rm c})\right]\right\} \, .
\label{nc}
\end{equation}
Here $z_0$ is some function of the initial deviations $(\delta
l_0, \, \delta\theta_0)$\,. The exact form of $z_0$ is
unessential. For an estimate, we notice that in the limit $K\gg 1$
the evolution of $\delta\theta_n$ may be approximated as
$(\delta\theta_0+\delta l_0/K \cos\theta_0)\, \prod_{k=0}^{n-1} \,
K\cos\theta_k $\,. Therefore,
\begin{eqnarray}
z_0           &               =           &                \ln
\left|\delta\theta_0+\delta l_0/K \cos\theta_0\right|
\nonumber\\
           &               \simeq           &                \ln \sqrt{\delta\theta_0^2 +\delta l_0^2/K^2} \,
. \label{z0}
\end{eqnarray}

Substituting   Eq.~(\ref{nc}) into $\exp (2i\omega n_{\rm c})$, we
obtain:
\begin{eqnarray}
e^{2i\omega\, n_{\rm c}}            &               =           &
\exp\left[-\frac{2i\omega\, z_0 }{\lambda (n_{\rm c})}\right]
\label{Wave1}\\
           &                          &                \times \exp\left\{-\frac{2i\omega} {\lambda
(n_c)}\sum_{j=0}^{n_{\rm c}-1} \left[\ln \left(1+\alpha|_j\right)-
\lambda
(n_{\rm c})\right]\right\} \nonumber\\
           &               \approx           &                \exp\left[-\frac{2i\omega\, z_0 }{\lambda
(n_{\rm c})}\right]\, \exp \left[-\frac{2\omega^2 \lambda_2
(n_{\rm c}) \, z_0}{\lambda^2 (n_{\rm c})}\right] \, , \nonumber
\end{eqnarray}
where in the second line we use the fact that $\omega \lambda_2
(n_{\rm c}) n_{\rm c}/\lambda^2 (n_{\rm c}) \ll 1$\,.  In the
limit $n_{\rm c}\gg 1$\,, $\lambda (n_{\rm c})\rightarrow
\lambda,\, \lambda_2 (n_{\rm c})\rightarrow \lambda_2$\,.
Moreover, from Eq.~(\ref{nc}) we see that $n_{\rm c}\rightarrow
t^{\rm C}\equiv -z_0/\lambda$\,. Thus, Eq.~(\ref{Wave1}) is
reduced to Eq.~(\ref{Gamma}).

\begin{figure}[h]
  \centerline{\epsfxsize=3in \epsfbox{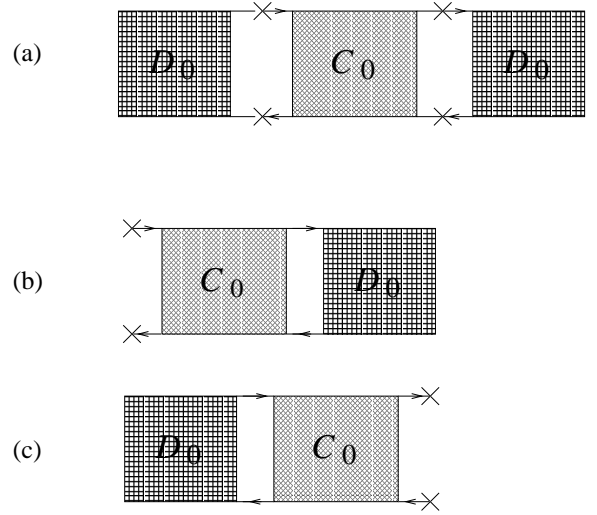}}
\caption{ Diagrams that lead to the interaction vertex with the
particle conservation law respected: (a) $\delta {\hat {\cal
D}}_1$, (b) $\delta {\hat {\cal D}}_2$, and (c) $\delta {\hat
{\cal D}}_3$.} \label{figoneloopdiagram}
\end{figure}

\section{The exact interaction vertex}
\label{WardCancellation}

In this Appendix we discuss the exact interaction vertex appearing
in the one--loop calculation. The general diagram, sketched in
Fig.~\ref{oneloopfig}, may be categorized into three classes,
Fig.~\ref{figoneloopdiagram} a-c. In order to make the formula
compact, let us write them down in the operator representation as:
\begin{eqnarray}
\delta {\hat {\cal D}}_1            &               =           &
\left( e^{i\omega}\, {\hat {\cal P}} {\hat {\cal D}}\right)\,
{\hat {\bf {\rm C}}} \, \left( e^{i\omega}\, {\hat {\cal P}} {\hat
{\cal D}}\right)
\, , \label{oneloopD1}\\
\delta {\hat {\cal D}}_2            &               =           &
{\hat {\cal P}}_V {\hat {\bf {\rm C}}} {\hat {\cal D}}
\, , \label{oneloopD2}\\
\delta {\hat {\cal D}}_3            &               =           &
{\hat {\cal D}} {\hat {\bf {\rm C}}} {\hat {\cal P}}_V\, .
\label{oneloopD3}
\end{eqnarray}

We then insert the identity \,:
\begin{equation}
e^{i\omega}\, {\hat {\cal P}}\, \equiv 1+\left(e^{i\omega}\, {\hat
{\cal P}}-1 \right) \label{Piden}
\end{equation}
into Eq.~(\ref{oneloopD1}) and rewrite $\delta {\hat {\cal D}}_1$
as
\begin{equation}
\delta {\hat {\cal D}}_1 \equiv {\hat {\cal D}}\, {\hat {\bf C}}\,
{\hat {\cal D}}+\delta {\hat {\cal D}}_1'+\delta {\hat {\cal D}}_4
\label{D1decomposition}
\end{equation}
with
\begin{eqnarray}
\delta {\hat {\cal D}}_1'            &               =           &
\left[\left(e^{i\omega}\, {\hat {\cal P}}-1 \right) \,{\hat {\cal
D}}\right]{\hat {\bf {\rm C}}} \, {\hat {\cal D}} +
 {\hat {\cal D}}\, {\hat {\bf {\rm C}}}\, \left[\left(e^{i\omega}\, {\hat {\cal
P}}-1
\right) \,{\hat {\cal D}}\right] \nonumber\\
\nonumber\\
\delta {\hat {\cal D}}_4            &               =           &
\left[\left(e^{i\omega}\, {\hat {\cal P}}-1 \right) \,{\hat {\cal
D}}\right]\, {\hat {\bf {\rm C}}}\, \left[\left(e^{i\omega}\,
{\hat {\cal P}}-1 \right) \,{\hat {\cal D}}\right]
 \, .
\label{D1D4}
\end{eqnarray}
For the first term in Eq.~(\ref{D1decomposition}), it was proved
by Altland \cite{Altland93} that it vanishes in the quantum limit.
In the next section, we show that this term is indeed pure
classical, and does not contribute to the quantum interference
correction. By the same token we show that  the loop expansion
does not violate the particle conservation law. That is the form
of the interaction vertex implies that the quantum corrections may
be reduced to the renormalization of the diffusion coefficient.

\subsection{Semiclassical analysis on ${\hat {\cal D}}\, {\hat {\bf C}}\,
{\hat {\cal D}}$} \label{wardiden}

Under appropriate approximation, it was proved that ${\hat {\cal
D}}\, {\hat {\bf C}}\, {\hat {\cal D}}$ is included into {\it
classical diffusive} propagator  \cite{Altland93}. Thus it does
not violate diffusion equation. Since we are interested in the
dynamics involving Ehrenfest time, the diffusive propagator can
not serve as a starting point of the formalism developed here. A
natural question is whether this conclusion may still be
applicable. Although it remains a challenge to prove at the
accurate level, here we present a physical interpretation. We
conclude that this term is pure classical, included in the exact
classical propagator--solution of the FPR equation.

For quantitative discussions below, let us denote the angular
momenta and angles, appearing in the retarded/advanced propagator
lines of Cooperon as $l_{1\pm}\, ,l_{2\pm}\, ,\cdots ,\, l_{n\pm}$
and $\theta_{1\pm}\, ,\theta_{2\pm}\, ,\cdots ,\,
\theta_{(n-1)\pm}$ following the forward time direction of the
retarded line (see Fig.~\ref{oneloopfig} for notations). Counting
from the most left hand side of the Cooperon and following the
backward time direction of the retarded line, we denote those as
$l_{0\pm}\, ,l_{-1\pm}\, ,l_{-2\pm}\, ,\cdots $, and
$\theta_{0\pm}\, ,\theta_{-1\pm}\, ,\theta_{-2\pm}\, ,\cdots $ .
Semiclassically, only the constraint below
\begin{equation}
|\theta_{1+}+\theta_{(n-1)-}|\sim
|\theta_{2+}+\theta_{(n-2)-}|\cdots |\theta_{(n-1)+}+\theta_{1-}|
\lesssim  \kbar \label{CooperonDefinition}
\end{equation}
is imposed.

According to the exact Eq.~(\ref{qEq}), at the boundary of
Cooperon and (left) Diffuson,
\begin{equation}
\left(\theta_{1+} - \theta_{1-}\right) = \left(\theta_{0+} -
\theta_{0-}\right) -\left(l_{1+} - \theta_{1-}\right) \, .
\label{qEqCoop}
\end{equation}
On the other hand, Eq.~(\ref{CooperonDefinition}) does not impose
any constraints on $\theta_{1+} - \theta_{1-}$. If it is order
$\kbar$, the usual Wigner transform may be performed (see
Eq.~(\ref{DWig})). That is to say, the left-most kicks of the
Cooperon may be incorporated into the left Diffuson.

This discussions remain applicable for all the successive pairs of
kicks of the Cooperon, until we meet some $k  >  1$, where
$\theta_{k+} - \theta_{k-} \sim 1$--a typical feature of the
Cooperon. This is equivalent to the validity of
Eq.~(\ref{qEqCoop}) with the right hand side of order $1$. In
fact, $\left(\theta_{0+} - \theta_{0-}\right)$ and $\left(l_{1+} -
\theta_{1-}\right)$ develop from the left (classical)
Diffuson--following Eqs.~(\ref{qEq}) and (\ref{mEq}). In the
Diffuson side, $\left(\theta_{k+} - \theta_{k-}\right)\,
(k=-1,-2,\cdots)$, $\left(l_{k+} - l_{k-}\right)\,
(k=0,-1,-2,\cdots)$ are of order $\kbar$. Moreover, any evolution
can last at most finite classical correlation time $\tau_c$. Thus,
$(\theta_{1+} - \theta_{1-})$ must be of order $\kbar$. Thus,
Eq.~(\ref{qEqCoop}) can not be satisfied if $(\theta_{1+} -
\theta_{1-}) \sim {\cal O} (1)$.

Thus we may conclude that ${\hat {\cal D}}\, {\hat {\bf C}}\,
{\hat {\cal D}}$ is pure classical. It characterizes the classical
probability of trajectories with a peculiar feature. Since
$|\theta_{1-}+\theta_{n+}| \lesssim \kbar \ll 1$, $\theta_{1+}
\sim \theta_{1-}\sim -\theta_{n+}$, the trajectory switches its
angle to the direction (almost) opposite to the initial.

\subsection{A cancellation mechanism}
\label{derivationCancell}

Having the general expression, Eq.~(\ref{1loopCC}) at hand, we
turn to show that $\delta {\hat {\cal D}}_1'$\,, $\delta {\hat
{\cal D}}_2$, $\delta {\hat {\cal D}}_3$ cancel each other at the
semiclassical level, namely:
\begin{equation}
\delta {\cal D}_1'+\delta {\cal D}_2+\delta {\cal D}_3 \equiv 0\,
. \label{cancellationD1D2D3}
\end{equation}
Indeed, employing  identity, Eq.~(\ref{leftrightPrela}), we
obtain:
\begin{eqnarray}
   &    &      \delta {\cal D}_1'\left(l,\theta;l',\theta'\right)
   \label{vertex11}\\
   &    =  &                {\hat {\cal
V}}\,  \bigg \{ {\cal D}_0\left(l,\theta;l''+\frac{\delta
l_2}{2},\theta''+\frac{\delta\theta_1}{2}\right)
  \Big [\left(e^{i\omega}\,
\overleftarrow{P}_T-1\right) \nonumber\\
   &       &    +\left(e^{i\omega}\,
\overrightarrow{P}-1\right)\Big ]   {\cal
D}_0\left(l''-\frac{\delta
l_2}{2},-\theta''+\frac{\delta\theta_1}{2};l',\theta'\right)\bigg
\} \, , \nonumber
\end{eqnarray}
\begin{eqnarray}
           &                          &                \delta {\cal D}_2 \left(l,\theta;l',\theta'\right)
\label{vertex2}\\
           &               =           &               {\hat {\cal V}}\,\bigg \{ 2\pi\kbar\,\delta
\left[l-\left(l''+\frac{\delta l_2}{2}\right)\right]\,  \delta
\left[\theta-\left(\theta''+\frac{\delta\theta_1}{2}\right)-l\right]
\nonumber\\
           &                          &                \times {\cal D}_0\left(l''-\frac{\delta
l_2}{2},-\theta''+\frac{\delta\theta_1}{2};l',\theta'\right)
\bigg\} \, , \nonumber
\end{eqnarray}
and
\begin{eqnarray}
\delta {\cal D}_3 \left(l,\theta;l',\theta'\right) = {\hat {\cal
V}}\, \bigg [ {\cal D}_0 \left(l,\theta;l''+\frac{\delta
l_2}{2},\theta''+\frac{\delta\theta_1}{2}\right) \nonumber\\
 \times   2\pi\kbar\, \delta \left(l''-\frac{\delta
l_2}{2}-l'\right) \, \delta
\left(-\theta''+\frac{\delta\theta_1}{2}-\theta'-l''-\frac{\delta
l_2}{2}\right) \bigg ]\, . \nonumber\\
\label{vertex3}
\end{eqnarray}
Since ${\cal D}_0$ is the solution of FPR equation, i.e.,
Eq.~(\ref{DysonCorrelator}), these three terms cancel each other.
Therefore we find that only $\delta {\cal D}_4$ leads to the
non--vanishing one-loop correction to the classical
density-density correlator.

\section{Aleiner-Larkin regularization in kicked rotor}
\label{ALRegularization}

It is easy to see that in the classical limit: $\kbar\rightarrow
0$,  the functional form of $\delta {\cal D}_1'$ and $\delta {\cal
D}_{2,3}$, i.e., Eqs.~(\ref{vertex11}), (\ref{vertex2}) and
(\ref{vertex3}) are identical to what have been found for
classical Lorentz gas (leaving the feature of standard map aside).
These terms cancel each other, leading to the absence of the weak
localization correction. Remarkably, this cancellation still
holds, even if the initial minimal wave packet is taken into
account. On the other hand, the weak localization correction {\it
does} exist, given by the exact vertex $\delta {\cal D}_4$, which
is absent in the expansion of ballistic supersymmetric $\sigma$
model (without a regularizer). In this way, one may wonder that an
appropriate regularization may lead to a physical description of
weak localization in semiclassical chaotic systems. This, indeed,
was developed by Aleiner and Larkin for ballistic electronic
problems \cite{AL96,AL97}. However, an important issue remains
open whether and to what extent the physical results depend on the
regularizer, rather than on the intrinsic quantum nature of the
problem.  The exact interaction vertex may serve as a testing
ground of this regularization.

In fact, the regularization introduced in
Refs.~\onlinecite{AL96,AL97} is also applicable for QKR. To this
end  we try to mimic  the ``Born impurity'' by modifying the
(quantum) free rotation operator ${\hat {\cal P}}_V$ in the
following way:
\begin{equation}
{\hat {\cal P}}_V \rightarrow {\hat {\cal P}}_V \, \exp \left(
\frac{i\, \delta {\hat S}}{\kbar} \right) \, . \label{impurity}
\end{equation}
Here $\delta {\hat S}$ is some stationary random perturbation that
commutes with ${\hat {\cal P}}_V$\,, i.e., $[\delta {\hat S},\,
{\hat {\cal P}}_V]=0$\,. Moreover, we assume that it is
short--ranged correlated  in the angular momentum space:
\begin{equation}
\langle \partial_l\, \delta S_l \,\, \partial_{l'}\, \delta
S_{l'}\rangle=\frac{2}{\tau_q}\,\delta_{l,\,l'} \, . \label{RP}
\end{equation}
In the classical limit, this additional stationary random
perturbations leads to the modification of the standard mapping in
the following way
\begin{eqnarray}
l_{n+1}           &               =           &               l_n+K\sin\theta _n,\nonumber\\
\theta _{n+1}           &               =           & \theta _n+
l_{n+1}+
\partial_{l_{n+1}}\, \delta S_{l_{n+1}} \, .
\label{StandardMModification}
\end{eqnarray}
Then following the same procedure as in Ref.~\onlinecite{AL96}, we
find the one-loop correction to the density-density correlator to
be
\begin{eqnarray}
           &                          &                \delta {\cal D} \left( l,\theta; l',\theta' \right)
\nonumber\\
           &               =           &                \frac{2}{\tau _q} \int \frac{dl_1\, d\theta
_1}{2\pi\kbar} {\cal C} \left(l_l,\theta _1; l_1,-\theta _1
\right) \label{DiRenormal}
\\
           &        \times          &               \left[ \frac{\partial}{\partial\theta _1}
e^{i\omega} \overrightarrow{P} {\cal D} \left(l,\theta; l_1,\theta
_1 \right) \right]\, \left[ \frac{\partial}{\partial\theta _1}
e^{i\omega} \overrightarrow{P}{\cal D} \left(l_1,-\theta _1;
l',\theta ' \right) \right] \nonumber
\end{eqnarray}
with the regularized Diffuson ${\cal D}$ and the Cooperon ${\cal
C}$ satisfying the following equation:
\begin{eqnarray}
           &                          &                \left[1-\left(1+\frac{1}{\tau_q}\, \frac{\partial
^2}{\partial \theta ^2}\right) \, e^{i\omega}
\overrightarrow{P}\right ] \left\{
\begin{array}{c}
  {\cal D} \left( l,\theta; l',\theta' \right)\\
  {\cal C} \left( l,\theta; l',\theta' \right) \\
\end{array}
\right\} \nonumber\\
           &               =           &                2\pi\kbar\, \delta\left(l-l'\right)\delta\left(\theta
-\theta'\right) \, . \label{DRenormal}
\end{eqnarray}
Eqs.~(\ref{DiRenormal}) and (\ref{DRenormal}) are fully analogous
to those found for ballistic electronic systems. In
Eq.~(\ref{DRenormal}), the regularizer of FRP equation mimics the
spread of minimal wave packet arising from {\it intrinsic} quantum
diffractions. Originally, it was expected \cite{AL96} that the
only physical effect arising from this regularizer is to determine
the Ehrenfest time since it smears the sharp classical propagator.
Indeed, one may further follow the procedure of
Ref.~\onlinecite{AL96} to calculate the weak--localization
correction to the diffusion constant. As a result, the functional
form of Eq.~(\ref{DDclKW}) is reproduced with $t_E$ acquiring
explicit $\tau_q$-dependence \cite{AL96}. At $(\lambda
\tau_q)^{-1} \sim \kbar/K$, the result thereby obtained are
identical to Eq.~(\ref{DDclKW}). This reflects a basic belief
previously anticipated \cite{AL96,VL03,TL03}. That is, a physical
regularizer strength must match the minimal quantum wave packet.

\end{appendix}


\begin{thebibliography}{}

\bibitem{Basko}D. M. Basko, M. A. Skvortsov, and V. E. Kravtsov, Phys. Rev. Lett. {\bf 90}, 096801 (2003);
V. E. Kravtsov, cond-mat/0312316.

\bibitem{Raizen95}F. L. Moore {\it et al.}, Phys. Rev. Lett. {\bf 75},
4598 (1995); C. F. Bharucha {\it et al.}, Phys. Rev. E {\bf
60},3881 (1999).

\bibitem{Raizen99}M. G. Raizen, Adv. At., Mol., Opt. Phys. {\bf
41}, 43 (1999).

\bibitem{Zhang04}C. Zhang, J. Liu, M. Raizen, and Q. Niu, Phys. Rev.
Lett. {\bf 92}, 054101 (2004).

\bibitem{Jacquod} O. A. Starykh, P. R. J. Jacquod, E. E. Narimanov, and A. D.
Stone, Phys. Rev. E {\bf 62}, 2078 (2000).

\bibitem{CCFI79}G. Casati, B. V. Chirikov, J. Ford, and F. M.
Izrailev, in {\it Stochastic Behavior of Classical and Quantum
Hamiltonian Systems}, Lecture Notes in Physics {\bf 93}, edited by
G. Casati and J. Ford, 334 (Springer, New York 1979).

\bibitem{CIS81}B. V. Chirikov, F. M. Izrailev, and D. L. Shepelyansky,
Sov. Sci. Rev. Sec. C {\bf 2}, 209 (1981).

\bibitem{Ammann}H. Ammann {\it et al.}, Phys. Rev. Lett. {\bf 80},
4111 (1998).

\bibitem{Chirikov79}B. V. Chirikov, Phys. Rep. {\bf 52}, 263
(1979).

\bibitem{LL}A. L. Lichtenberg and M. A. Lieberman, {\it Regular and Chaotic
Dynamics} (Springer-Verlag, Berlin 1991).

\bibitem{Rec81}A. B. Rechester and R. B. White, Phys. Rev. Lett. {\bf
44}, 1586 (1980); A. B. Rechester, M. N. Rosenbluth, and R. B.
White, Phys. Rev. A {\bf 23}, 2664 (1981).

\bibitem{KFA00}M. Khodas and S. Fishman, Phys. Rev. Lett. {\bf 84},
2837 (2000); Erratum, {\it ibid.} {\bf 84}, 5918 (2000); M.
Khodas, S. Fishman, and O. Agam, Phys. Rev. E {\bf 62}, 4769
(2000).

\bibitem{FGP82}S. Fishman, D. R. Grempel, and R. E. Prange, Phys. Rev.
Lett. {\bf 49}, 509 (1982).

\bibitem{FGP84}S. Fishman, D. R. Grempel, and R. E. Prange, Phys. Rev. A
{\bf 29}, 1639 (1984).

\bibitem{BS92}R. Bl{\"u}mel and U. Smilansky, Phys. Rev.
Lett. {\bf 69}, 217 (1992).

\bibitem{LR85}P. A. Lee and T. V. Ramakrishnan, Rev. Mod. Phys. {\bf
57}, 287 (1985).

\bibitem{GLKh79}L. P. Gorkov, A. I. Larkin, and D. E. Khmelnitskii,
Pis'ma Zh. Eksp. Teor. Fiz. {\bf 30}, 248 (1979) [JETP Lett. {\bf
30} , 248 (1979)].

\bibitem{Altland93}A. Altland, Phys. Rev. Lett. {\bf 71}, 69 (1993).

\bibitem{AZ96}A. Altland and M. R. Zirnbauer, Phys. Rev. Lett. {\bf
77}, 4536 (1996).

\bibitem{Efetov97} K. Efetov, {\it Supersymmetry in disorder and
chaos} (Cambridge University Press, UK 1997).

\bibitem{LO68}A. I. Larkin and Yu. N. Ovchinnikov, Zh. Eksp. Teor.
Fiz. {\bf 55}, 2262 (1968) [Sov. Phys. JETP {\bf 28}, 1200
(1969)].

\bibitem{BZ78}G. P. Berman and G. M. Zaslavsky, Physica A {\bf 91}, 450
(1978); G. M. Zaslavsky, Phys. Rep. {\bf 80}, 157 (1981).

\bibitem{Izrailev90}For a review, see F. M. Izrailev, Phys. Rep. {\bf 196}, 299
(1990).

\bibitem{Beenakker} Ph. Jacquod and E. V. Sukhorukov, Phys. Rev.
Lett. {\bf 92}, 116801 (2004); J. Tworzydlo, A. Tajic, and C. W.
J. Beenakker, Phys. Rev. B {\bf 69}, 165318 (2004),
cond-mat/0405122.

\bibitem{Beenakker03} M. C. Goorden, Ph. Jacquod, and C. W. J.
Beenakker, Phys. Rev. B {\bf 68}, 220501(R) (2003).

\bibitem{TTSB03} J. Tworzydlo, A. Tajic, H. Schomerus, and C. W. J. Beenakker,
Phys. Rev. B {\bf 68}, 115313 (2003); P. G. Silvestrov, M. C.
Goorden, and C. W. J. Beenakker, Phys. Rev. Lett. {\bf 90}, 116801
(2003).

\bibitem{AL96}I. L Aleiner and A. I. Larkin, Phys. Rev. B {\bf 54},
14423 (1996).

\bibitem{VL03} M. G. Vavilov and A. I. Larkin, Phys. Rev. B {\bf 67}, 115335
(2003).

\bibitem{TKL04} C. Tian, A. Kamenev, and A. Larkin, Phys. Rev.
Lett. {\bf 93}, 124101 (2004).

\bibitem{MA04} J. M{\"u}ler and A. Altland, nlin.CD/0411055.

\bibitem{She87}D. L. Shepelyansky, Physica D {\bf 28}, 103 (1987).

\bibitem{Raizen00}D. A. Steck, V. Milner, W. H. Oskay, and M. G.
Raizen, Phys. Rev. E {\bf 62}, 3461 (2000).

\bibitem{kbarexp}P. H. Jones {\it et. al.}, physics/0405046; P. H. Jones {\it et. al.},
quant-ph/0309149.

\bibitem{MRBF04}S. Montangero, A. Romito, G. Benenti, and R.
Fazio, cond-mat/0407274.

\bibitem{IGSGZ02} A. Isacsson, L. Y. Gorelik, R. I. Shekhter, Y. M. Galperin, and M. Jonson,
Phys. Rev. Lett. {\bf 89}, 277002 (2002);
L. Y. Gorelik, A. Isacsson, Y. M. Galperin, R. I. Shekhter, and M.
Jonson,
Nature {\bf 411},454 (2001).

\bibitem{Hikami81}S. Hikami, Phys. Rev. B {\bf 24}, 2671 (1981).

\bibitem{Zirnbauer99} M. R. Zirnbauer, in {\it Supersymmetry and Trace
Formulae, Chaos and Disorder}, edited by I. V. Lerner, J. P.
Keating and D. E. Khmelnitskii (Kluwer Academic/Plenum, New York
1999).

\bibitem{Zaslavsky97}See, e.g., G. M. Zaslavsky, M. Edelman, and
B. A. Niyazov, Chaos {\bf 7}, 159 (1997).

\bibitem{Izrailev86} F. M. Izrailev, Phys. Rev.
Lett. {\bf 56}, 541 (1986).

\bibitem{TBS93} M. Thaha, R. Bl{\"u}mel, and U. Smilansky, Phys. Rev.
E {\bf 48}, 1764 (1993).

\bibitem{TTB04} T. Tworzydlo, A. Tajic, and C. W. J. Beenakker,
cond-mat/0405122.

\bibitem{SLA98}R. A. Smith, I. V. Lerner, and B. L. Altshuler, Phys. Rev.
B {\bf 58}, 10343 (1998).

\bibitem{WLS99}R. S. Whitney, I. V. Lerner, and R. A. Smith, Waves
Random Media {\bf 9}, 179 (1999).

\bibitem{TL03}C. Tian and A. I. Larkin, Phys. Rev. B {\bf 70}, 035305 (2004).

\bibitem{Chu} For a review see e.g. S. Chu, Science {\bf 253}, 861
(1991); W. D. Phillips, Rev. Mod. Phys. {\bf 70}, 721 (1998); C.
E. Wieman, D. E. Pritchard, and D. J. Wineland, Rev. Mod. Phys.
{\bf 71}, S253 (1999).


\bibitem{GSZ}R. Graham, M. Schlautmann, and P. Zoller, Phys. Rev. A {\bf 45}, R19 (1992).

\bibitem{Raizen94}F. L. Moore {\it et al.}, Phys. Rev. Lett. {\bf 73},
2974 (1994); J. C. Robinson {\it et al.}, Phys. Rev. Lett. {\bf
74}, 3963 (1995).


%


\bibitem{footnoteKP} In the conventional QKR is periodic, i.e.
$\Psi(\theta+2\pi)=\Psi(\theta)$, leading to the discrete angular
momentum: $l=n\kbar$ with $n$ integer. Periodically kicked cold
atomic gases differs from it, with the boundary condition replaced
by: $\Psi(\theta+2\pi)=\Psi(\theta)e^{i\theta l_0}$, where $l_0\in
[0,1]$ is a fractional part of an atom's momentum in units $2\hbar
k_L$. It may be eliminated by the gauge transformation that
introduces an Aharonov-Bohm flux  into the kinetic energy $(\hat
l-l_0)^2/2$. Such flux does not affect dynamical localization,
though quantum resonances \cite{Izrailev90} is eliminated by
averaging over $l_0$.

\bibitem{TKL04a} C. Tian, A. Kamenev, and A. I. Larkin,
Bull. Am. Phys. Soc. {\bf 49}, 482 (2004).

\bibitem{CDH98}E. Chow, P. Delsing, and D. B. Haviland,
Phys. Rev. Lett. {\bf 81}, 204 (1998).

\bibitem{Haviland1} D. B. Haviland and P. Delsing,
Phys. Rev. B {\bf 54}, R6857 (1996).

\bibitem{Haviland2}M. Watanabe and D. B. Haviland,
in {\it Studies of High Temperature Superconductors (Advances in
Research and Applications)}, Vol. 43, edited by A. Narlikar (Nova
Science Publishers, New York, 2002).

\bibitem{Cohen91}D. Cohen, Phys. Rev. A {\bf 44}, 2292 (1991).

\bibitem{Milner00} V. Milner {\it et. al.}, Phys. Rev. E {\bf 61}, 7223
(2000).

\bibitem{Raizen98} B. G. Klappauf {\it et. al.}, Phys. Rev. Lett. {\bf 81},
1203 (1998).

\bibitem{DP02}A. J. Daley and A. S. Parkins, Phys. Rev. E {\bf 66}, 056210 (2002);
G. Duffy {\it et. al.}, cond-mat/0401346.

\bibitem{SET03} K. B. Efetov, G. Schwiete, and K. Takahashi, Phys. Rev. Lett. {\bf 92}, 026807
(2004).

\bibitem{AL97} I. L. Aleiner and A. I. Larkin, Phys. Rev. E {\bf 55}, R1243 (1997).

\end{thebibliography}
\end{document}